\documentclass{vldb}
\usepackage{graphicx}
\usepackage{balance}  
\usepackage{comment} 
\usepackage{lipsum} 
\usepackage{fullpage} 
\usepackage{amsfonts}
\usepackage{bm}
\usepackage{amsmath, amssymb}
\newtheorem{definition}{Definition}
\usepackage[lined, ruled,vlined]{algorithm2e}
\usepackage{subfig}
\usepackage{hhline}
\usepackage{color}
\usepackage{mathtools}
\usepackage[para,online,flushleft]{threeparttable}
\usepackage{booktabs}
\usepackage{pifont}
\newcommand{\xmark}{\ding{55}}

\usepackage{etoolbox}
\AtBeginEnvironment{align}{\setcounter{equation}{0}}
\renewcommand{\vec}[1]{\mathbf{#1}}

\begin{document}


\title{A Jointly Learned Context-Aware Place of Interest Embedding for Trip Recommendations}



%
%
%
%

\numberofauthors{3} 
\author{%
  {Jiayuan He, Jianzhong Qi, Kotagiri Ramamohanarao}%
  \vspace{1.6mm}\\
  \fontsize{10}{10}\selectfont\rmfamily\itshape
  School of Computing and Information Systems, The University of Melbourne, Australia\\
  \fontsize{9}{9}\selectfont\ttfamily\upshape
  hjhe@student.unimelb.edu.au, $\{$jianzhong.qi,kotagiri$\}$@unimelb.edu.au 
}
\date{30 July 1999}

\maketitle

\begin{abstract}
Trip recommendation is an important location-based service that helps 
relieve users from the time and efforts for trip planning. 
It aims to recommend a sequence of places of interest (POIs) for a user to visit that maximizes the user's satisfaction. When adding a POI to a recommended trip, 
it is essential to understand the context of the recommendation, including 
the POI popularity, other POIs co-occurring 
in the trip, and the preferences of the user. 
These contextual factors are learned separately in existing studies, while in reality, 
they impact jointly on a user's choice of a POI to visit. 
In this study, we propose a POI embedding model to jointly learn the impact of these contextual factors. 
We call the learned POI embedding a context-aware POI embedding. 
To showcase the effectiveness of this embedding, we apply it to generate trip recommendations given a user and a time budget. 
We propose two trip recommendation algorithms  based on our context-aware POI embedding. 
The first algorithm finds the exact optimal trip by transforming and solving the trip recommendation problem 
as an integer linear programming problem. To achieve a high computation efficiency, 
the second algorithm finds a heuristically optimal trip based on  adaptive large neighborhood search. 
We perform extensive experiments on real datasets. The results show that our proposed algorithms consistently outperform 
state-of-the-art algorithms in trip recommendation quality, 
 with an advantage of up to $43\%$ in F$_1$-score.

\end{abstract}

\section{Introduction}\label{sec:introduction}
Tourism is  one of the most profitable and fast-growing economic sectors in the world. In 2017, the tourism industry contributed more than 8.27 trillion U.S. dollars to global economy. 
The massive scale of the tourism industry calls for  more intelligent  services to improve user experiences and reduce labor costs of the industry. 
Trip recommendation is one of such services. Trip recommendation aims to recommend a sequence of \emph{places of interest} (POIs) for a user to vist
to maximize the user's satisfaction. Such a service benefits users by relieving them from the time and efforts for trip planning, which in return 
further boosts the tourism industry.

\begin{figure}
\centering
\includegraphics[width = 3.2 in]{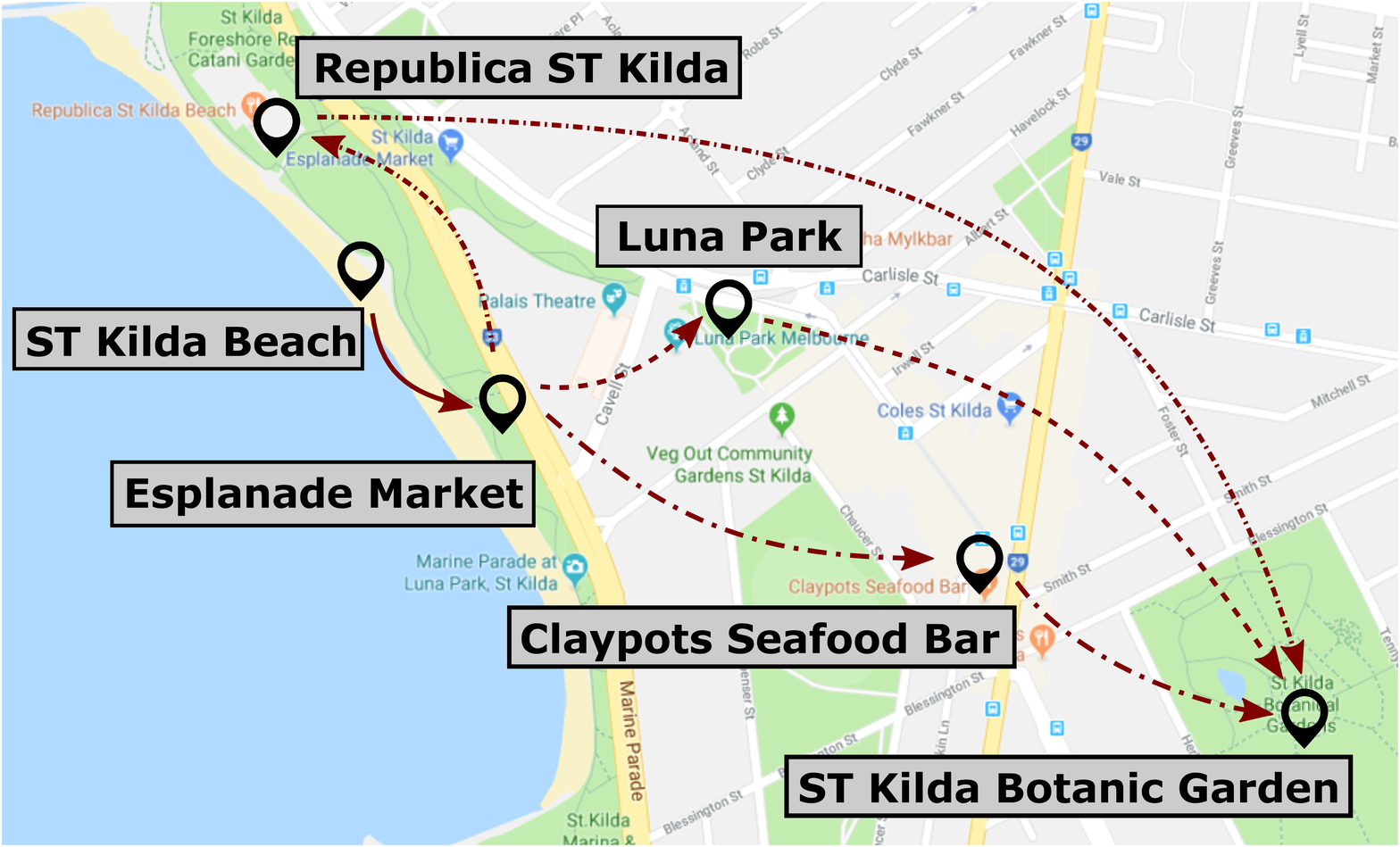}
\vspace{-5mm}
\caption{Impact of co-occurring POIs}
\label{fig:map}
\vspace{-5mm}
\end{figure}

%


Most existing studies on trip recommendations consider POI popularities or user preferences towards the POIs when making recommendations~\cite{cheng2011personalized,lim2015personalized}. 
Several recent studies~\cite{chen2016learning,rakesh2017probabilistic} consider the last POI visited when recommending the next POI to visit. 
These studies do not model the following two characteristics that we observe from real-world user trips (detailed in Section~\ref{sec:empirical}).
(i) A POI to be recommended is impacted not only by the last POI visited but also all other POIs co-occurring in the same trip. 
For example, in Fig.~\ref{fig:map}, a user has just visited ``ST Kilda Beach'' and ``Esplanade Market''. She may be tired after the long walk along the beach and the market. Thus, compared with ``Luna Park'' which is a theme park nearby, the user may prefer a restaurant (e.g., ``Republica ST Kilda'' or ``Claypots Seafood Bar'') to get some rest and food. 
The user plans to visit ``Botanic Garden'' later on. Thus, she decides to visit ``Claypots Seafood Bar'' since it is on the way from the beach to the garden. 
Here, the visit to ``Claypots Seafood Bar'' is impacted by the visits of not only ``Esplanade Market'' but also ``ST Kilda Beach'' and ``Botanic Garden.'' 
(ii) POI popularities, user preferences, and co-occurring POIs together impact the POIs to be recommended in a trip. In the example above, 
there can be many restaurants on the way to ``Botanic Garden.'' The choice of  ``Claypots Seafood Bar'' can be impacted by not only ``Botanic Garden'' but also 
the fact that the user is a seafood lover and that ``Claypots Seafood Bar'' is highly rated by other users. 
Most existing models~\cite{liu2016exploring, rakesh2017probabilistic}  learn the impact of each factor separately and simply combine them by linear summation, which may not reflect the joint impact accurately.

In this study, we model the two  observations above with a \emph{context-aware POI embedding model} to jointly learn the impact of POI popularities, user preferences, and co-occurring POIs. 
We start with modeling the impact of co-occurring POIs. 
Existing studies model the impact of the last POI with a first-order Markov model~\cite{chen2016learning, kurashima2010travel, rakesh2017probabilistic}. Such a model requires a large volume of data 
to learn the impact between every pair of adjacent POIs. However, real-world POI visits are sparse and highly skewed. Many POIs may not be adjacent in any trip and their impacts cannot be learned. 
Extending such a model to multiple co-occurring POIs requires a higher-order Markov model, which suffers further from the data sparsity limitation. 

We address the above data sparsity limitation 
by embedding the POIs into a space where POIs that co-occur frequently are close to each other. 
This is done based on our observation that a trip can be seen as a ``sentence'' where each POI visit is a ``word.'' 
The occurrence of a POI in a trip is determined by all the co-occurring words (POIs) in the same sentence (trip). 
This enables us to learn a POI embedding similar to the Word2Vec model~\cite{mikolov2013distributed} that embeds words into a space where words with a similar context are close to each other.

To further incorporate the impact of user preferences into the embedding, we project users into the same latent space of the POIs, 
where the preferences of each user is modeled by the proximity between the user and the POIs. 
We also extend the embedding of each POI by adding a dimension (a bias term) to represent the POI popularity. 
We jointly learn the embeddings of users and POIs via \emph{Bayesian Pairwise Ranking}~\cite{rendle2009bpr}.

To showcase the effectiveness of our proposed context-aware POI embedding, we apply it to a trip recommendation problem name TripRec where a user and her time budget is given. 
We propose two algorithms for the problem. The first algorithm, \emph{C-ILP}, models the trip recommendation problem as an integer linear programming problem. It solves the problem 
with an  integer linear programming technique~\cite{berkelaar2004lpsolve}.
C-ILP offers exact optimal trips, but it may be less efficient for large time budgets. 
To achieve a high efficiency, we further propose a heuristic algorithm named \emph{C-ALNS} based on the \emph{adaptive large neighborhood search} (ALNS) technique~\cite{ropke2006adaptive}. 
C-ALNS starts with a set of initial trips and optimizes them iteratively by replacing POIs in the trips with unvisited POIs that do not break the user time budget. 
We use the POI-user proximity computed by our context-aware POI embedding to guide
the optimization process of C-ALNS. This leads to high quality trips with low computational costs.  

This paper makes the following contributions:
\begin{enumerate}

\item We analyze real-world POI check-in data to show the impact of co-occurring POIs  
and the joint impact of contextual factors on users' POI visits.

\item
We propose a novel model to learn the impact of all co-occurring POIs rather than just the last POI in the same trip. 
We further propose a context-aware POI embedding model to jointly learn the impact of POI popularities, co-occurring POIs, and user preferences on POI visits.

\item
We propose two algorithms C-ILP and C-ALNS to generate trip recommendations based on our context-aware POI embedding model. 
C-ILP transforms trip recommendation to an integer linear programming problem and provides exact optimal trips. 
C-ALNS adapts the approximate large neighborhood search technique and provides heuristically optimal trips close to the exact optimal trips with a high efficiency.

\item
We conduct extensive experiments on real datasets. The  results show that our proposed algorithms outperform state-of-the-art algorithms consistently 
in the quality of the trips recommended as measured by the F$_1$-score. Further, our heuristic algorithm C-ALNS produces trip recommendations 
that differ in accuracy from those of C-ILP by only 0.2\% while reducing the running time by 99.4\%.

\end{enumerate}

The rest of this paper is structured as follows. Section~\ref{sec:related} reviews related studies. 
Section~\ref{sec:empirical} presents an empirical analysis on real-world check-in datasets to show the impact factors of POI visits. 
Section~\ref{sec:problem} formulates the problem studied. Section~\ref{sec:model} 
details our POI embedding model, and Section~\ref{sec:algorithms} details our trip recommendation algorithms based on the model. 
Section~\ref{sec:experiments} reports experiment results. Section~\ref{sec:conclusions} concludes the paper.

\section{Related work}
\label{sec:related}
We compute POI embeddings to enable predicting POI sequences (trips) to be recommended to users. 
We review inference models for predicting a POI to be recommended in Section~\ref{sec:lit_model}. 
We review trip generation algorithms based on these models in Section~\ref{sec:lit_gen}.

\subsection{POI Inference Model\label{sec:lit_model}} 
Most existing inference models for trip recommendations assume POIs to be independent from each other, i.e., the probability of a POI to be recommended is independent from that of any other POIs~\cite{brilhante2013shall,ge2011cost,lim2015personalized,  wang2016improving}. For example, Brilhante et al.~\cite{ge2011cost} assume that the probability 
of a POI to be recommended is a weighted sum of a popularity score and a user interest score, where the user interest score is computed via user-based collaborative filtering.
Assuming independence between POIs loses the POI co-occurrence relationships, we do not discuss studies based on this assumption further. 


Kurashima et al.~\cite{kurashima2010travel} propose the first work that captures POI dependency. They use the Markov model to capture the dependence of a POI $l_{i+1}$ on its preceding POI $l_{i}$ in a trip as the transition probability from $l_i$ to $l_{i+1}$. 
Rakesh et al.~\cite{rakesh2017probabilistic} also assume that each POI visit depends on its preceding POI. They unify such dependency with other factors (e.g., POI popularities) into a latent topic model. 
The model  represents each user's preference as a probability distribution over a set of latent topics. Each latent topic in turn is represented as a probability distribution over POIs. To capture the dependency between consecutive POI visits, they assume that the probability distribution of a latent topic changes with the preceding POI visit. 
Both these two studies~\cite{kurashima2010travel,rakesh2017probabilistic}
suffer from the data sparsity problem as they aim to learn the transition probability between any two adjacent POIs.
For many POI pairs,  there may not be enough transitions between them observed in real-world POI check-in data, 
because check-ins at POIs are highly skewed towards the most popular POIs. 
This may lead to unreliable transition probabilities and suboptimal trip recommendations. 
Our model does not require the POIs to be adjacent to learn their transition probability.
This helps alleviate the data sparsity problem, which leads to improved trip recommendations.

Chen et al.~\cite{chen2016learning} also use the Markov model to capture the dependence between POIs. To overcome the data sparsity problem, they factorize the transition probability between two POIs as the product of the pairwise transition probabilities w.r.t. five pre-defined features: POI category, neighborhood (geographical POI cluster membership), popularity (number of distinct visitors), visit counts (total number of check-ins), and average visit duration. 
These five features can be considered as an embedding of a POI. Such an embedding is manually designed rather than being learned from the data. It may not reflect the salient features of a POI. 

POI dependency is also considered in POI recommendations~\cite{cheng2012fused, liu2017experimental,miao2016s,ye2013s}, 
which aim to recommend an individual POI instead of a POI sequence. Such studies do not need 
to consider the dependence among the POIs in a trip. 
For example, Ye et al.~\cite{ye2013s} propose a \emph{hidden Markov model} (HMM) for POI recommendation. This model captures the transition probabilities between POIs assuming the POI categories as the hidden states. To recommend a POI, Ye et al. first predict the POI category of the user's next check-in. Then, they predict a POI according to the user's preferences over POIs within the  predicted POI category. 
Feng et al.~\cite{feng2015personalized} project POIs into a latent space where the pairwise POI distance represents the transition probabilities between POIs. Liu et al.~\cite{liu2016exploring} also use a latent space for POI recommendations. They first learn the latent vectors of POIs to capture the dependence between POIs. Then, they fix the POI vectors and learn the latent vectors of users from the user-POI interactions. 
These studies differ from ours in three aspects: (i) Their models learn the impact of  POIs  and the impact of user preferences independently, while our model learns the impact of the two factors jointly,  
which better captures the data characteristics and leads to an improved trip recommendation quality as shown in our experimental study. (ii) Their models focus on user preferences  and do not consider the impact of POI popularities, while ours take both into consideration. (iii) These studies do not consider constraints such as time budgets while ours does.

\subsection{Trip Generation\label{sec:lit_gen}} 

Trip recommendation aims to generate a trip, i.e., a sequence of POIs, that meets user constraints and maximizes user satisfaction. 
Different user constraints and user satisfaction formulation differentiate  trip recommendation studies. 
For example, Brilhante et al.~\cite{brilhante2013shall} consider a user given  time budget.  
They partition historical trips into segments each of which is associated with a time cost. Then, they reduce the trip recommendation problem  to a \emph{generalized maximum coverage} (GMC) problem that finds trip segments whose time costs together do not exceed the user time budget, while 
a user satisfaction function is maximized. 
Gionis et al.~\cite{gionis2014customized} assume a given sequence of POI categories and  the minimum and maximum numbers of POIs to recommend for each category. 
They use  dynamic programming  to compute trip recommendations. 
Lim et al.~\cite{lim2015personalized} formulate trip recommendation as an \emph{orienteering problem} that recommends a trip given a starting POI, an ending POI, 
and a time budget. They adopt the \emph{lpsolve} linear programming package~\cite{berkelaar2004lpsolve} to solve the problem. 
To showcase the applicability of our context-aware POI embedding model, we apply it to the trip recommendation problem studied by Lim et al.~\cite{lim2015personalized}. 
As we consider the joint impact of contextual factors,  
our user satisfaction formulation becomes nonlinear, which cannot be optimized by Lim et al.'s approach. 

Among the studies that consider POI dependency,  
Hsieh et al.~\cite{hsieh2014mining} and Rakesh et al.~\cite{rakesh2017probabilistic} assume a given 
 starting POI $l_s$, a given time budget $t_q$, and a given time buffer $b$. 
 They build a trip recommendation by starting from $l_s$ and progressively adding more POIs to the trip until the trip time reaches $t_q-b$. 
 They repeatedly add the unvisited POI that has the highest transition probability from the last POI in the trip. 
 As discussed earlier, their transition probabilities depend only on the last POI but not any other co-occurring POIs. 
 Chen et al.~\cite{chen2016learning} assume given starting and ending POIs and a time budget. 
 They formulate trip recommendation as an orienteering problem in a directed graph, where every vertex represents a POI and the weight of an edge represents the transition probability from its source vertex to its end vertex. Our trip recommendation problem share similar settings.  However,  Chen et al.'s algorithm does not apply to our problem as we not only consider the transition probabilities between adjacent POIs
  but also the impact of all POIs in a trip.

\section{Observations on POI Check-ins}
\label{sec:empirical}

We start with an empirical study on real-world POI check-in data to observe 
users' check-in patterns.  
We aim to answer the following three questions: (1) Are users' check-ins at a POI  impacted by  other POIs  co-occurring in the same trip? 
(2)~Are users' check-ins at a POI impacted by (other users') historical check-ins at the POI, i.e., the popularity of the POI?
(3)~Are the impact of co-occurring POIs and the impact of POI popularity independent from each other? 

\begin{table}
\renewcommand*{\arraystretch}{1.0}
\centering
\small
\begin{scriptsize}
\caption{Dataset Statistics\label{tab:datasets}}
\vspace{-3mm}
\begin{threeparttable}
\begin{tabular}{lllll}
\toprule[1pt]
Dataset & \#users & \#POI visits & \#trips & POIs/trip\\ \midrule[1pt]
Edinburgh & 82,060 & 33,944 & 5,028 & 6.75 \\\midrule
Glasgow & 29,019 & 11,434  & 2,227 & 5.13 \\\midrule
Osaka & 392,420  & 7,747 & 1,115 &  6.95 \\\midrule
Toronto & 157,505  & 39,419 & 6,057 & 6.51   \\\bottomrule[1pt]
\end{tabular}
\vspace{-4mm}
\end{threeparttable}
\end{scriptsize}
\end{table}

We analyze four real check-in datasets used in trip recommendation studies~\cite{chen2016learning,lim2015personalized}.
These four datasets are extracted from the Yahoo!Flickr Creative Commons 100M (YFCC100M) dataset~\cite{thomee2016yfcc100m}. 
They contain check-ins in the cities of Edinburgh, Glasgow, Osaka, and Toronto respectively. 
Table~\ref{tab:datasets} summarizes the statistics of the four datasets. 
For example, the Edinburgh dataset contains 33,944 POI visits from 82,060 users (consecutive check-ins at the same POI is counted as a POI visit).
The POI visits form 5,028 different trips, i.e., sequences of POI visits by the same user within 
an eight-hour period. There are 6.75 POI visits per trip on average.

\textbf{Impact of co-occurring POIs.} 
To verify the impact of co-occurring POIs, for each POI $l$, 
we compute the frequency distribution of the co-occurring POIs of $l$. 
If such frequency distributions of different POIs are different, then the POIs can be distinguished by such  
frequency distributions, and a POI visit can be determined by visits to the co-occurring POIs. 
This verifies the impact of co-occurring POIs.

\begin{figure}[h]
\vspace{-5mm}
\centering
\subfloat[Independent POI pairs]{\includegraphics[width = 1.5 in]{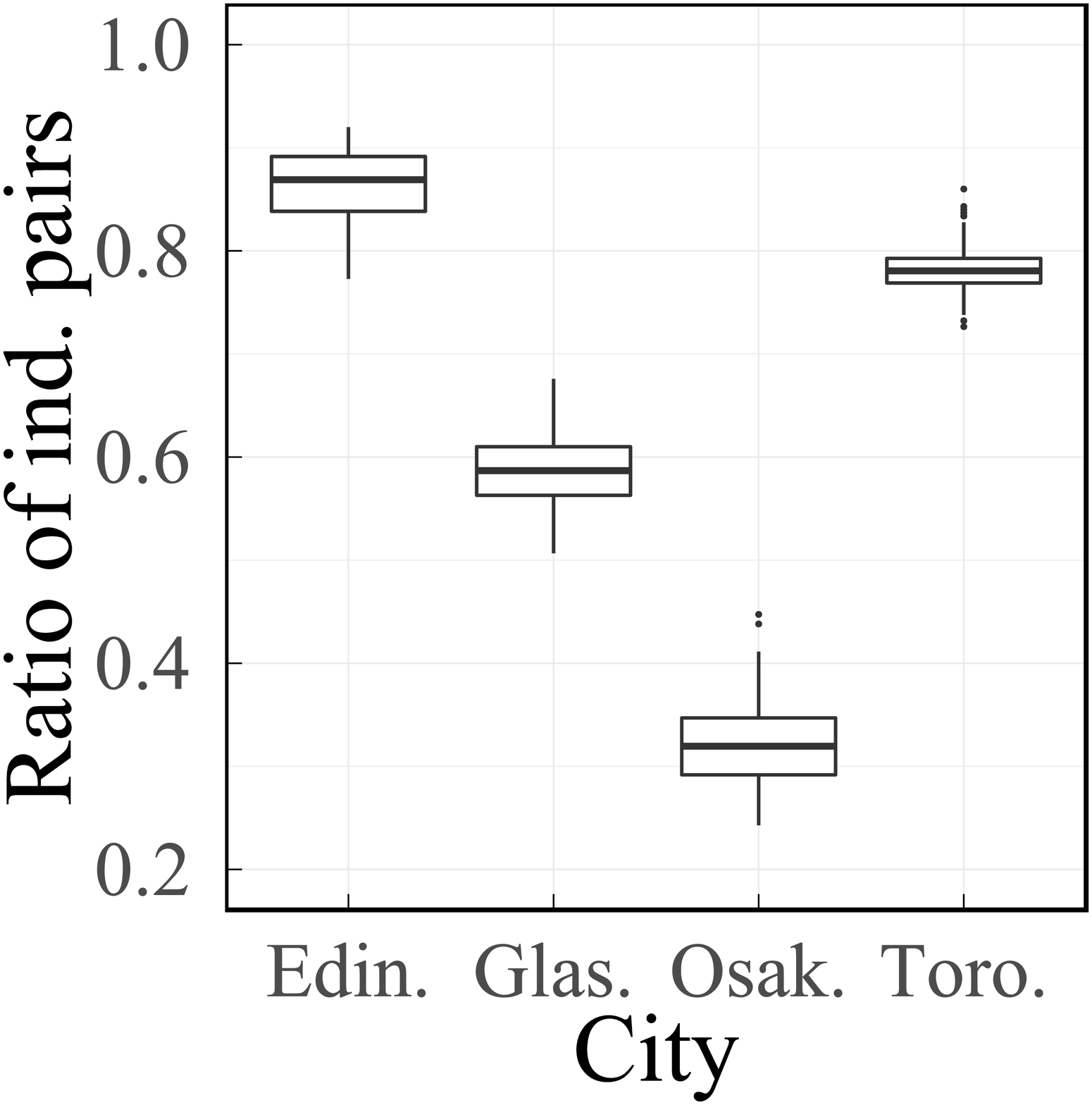}\label{fig:analysis_contextual}}\hspace{1mm}
\subfloat[Impacted users]{\includegraphics[width = 1.5 in]{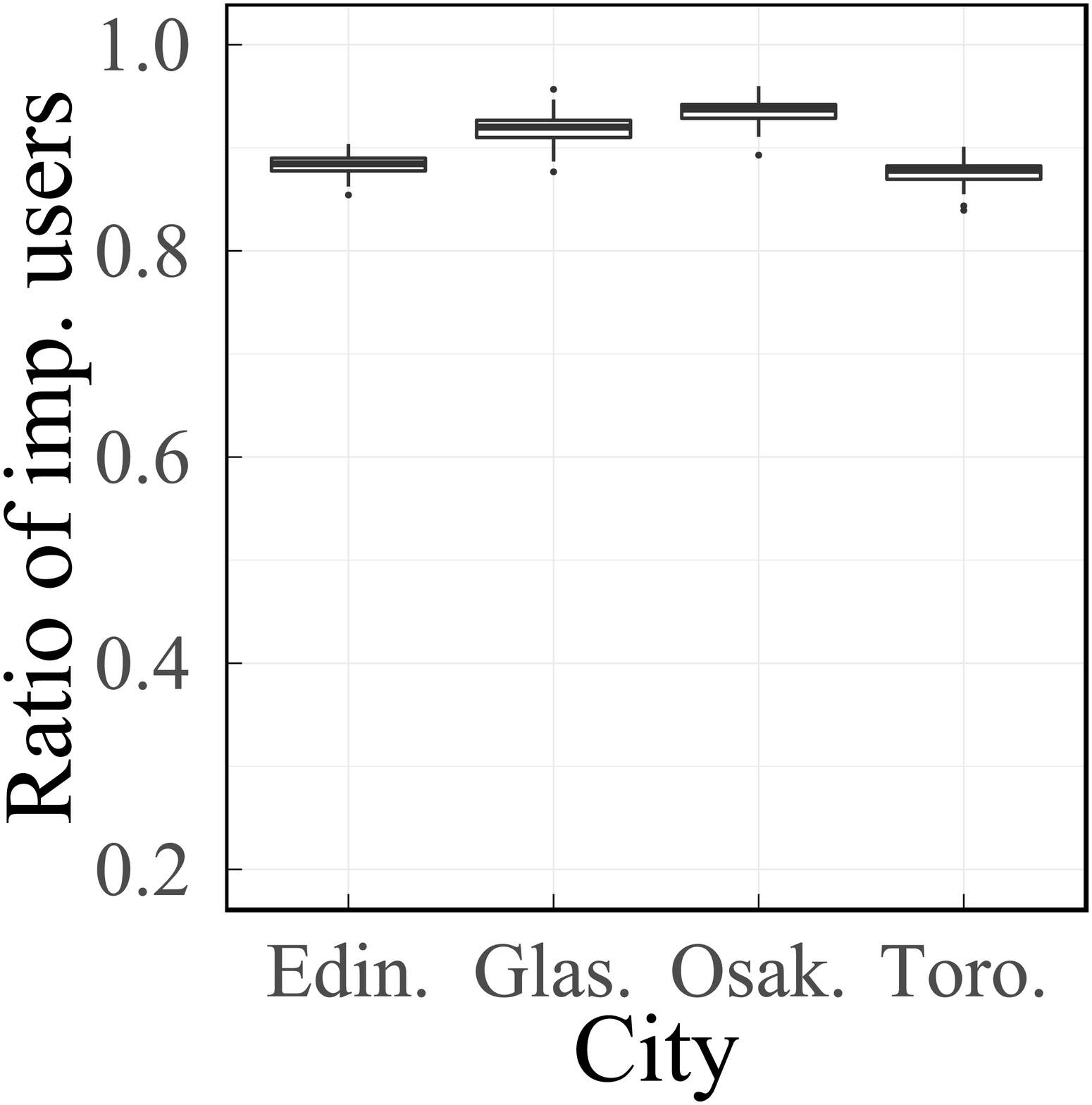}\label{fig:analysis_pop}}
\vspace{-2mm}
\caption{Observations on POI check-ins}
\label{fig:emp_analysis}
\vspace{-3mm}
\end{figure}

For each dataset, we perform a hypothesis test on whether 
two POIs have different frequency distributions of co-occurring POIs as follows. 
We randomly sample $50\%$ of the trips. 
From the sampled dataset, for each POI $l$, we compute an $|\mathcal{L}|$-dimensional distribution named the \emph{co-occurrence distribution}, 
where $\mathcal{L}$ is the set of all POIs in the dataset, and  dimension $i$ represents the normalized frequency of POI $l_i$ occurring in the same trip as $l$.
We perform a \emph{chi-square two sample test} for each pair of POIs on their co-occurrence distributions, 
where the null hypothesis is that ``the two distributions conform the same underlying distribution'' and the significance level is 0.05. 
If the hypothesis is rejected, we say that the two POIs form an \emph{independent POI pair}. 
We generate 100 sample datasets and report the average ratio of  independent POI pairs over all POI pairs. 
Figure~\ref{fig:analysis_contextual} shows the result, where each gray dot represents the ratio of a sample dataset, and the rectangles denote the 25 percentile, median, and 75 percentile. On average, independent POI pairs take up as least 32.5 (Osaka) and up to 87.5\% (Edinburgh) of all POI pairs. 
This means that a non-trivial portion of POIs have different co-occurrence distributions, which confirms the impact of co-occurring POIs.

\textbf{Impact of POI popularity.} 
POI popularity is commonly perceived to have a major impact on POI visits~\cite{ge2011cost}.
We add further evidence to this perception. 
For each city, we randomly split its dataset into two subsets, each of which consists of the POI visits of half of the users. 
We use one of the subsets as a \emph{historical dataset}, from which we compute a rank list of the POIs in $\mathcal{L}$ by their number of visits in the historical dataset. 
A POI with more visits ranks higher and is considered to be more popular. 
We use the other subset as a \emph{testing dataset}. For each user $u$ in the testing dataset, we test whether she visits the popular POIs in $\mathcal{L}$ 
more often than the less popular POIs. We compute the average rank of her visited POIs. If the average rank is higher than $|\mathcal{L}|/2$,
we consider $u$ to be an \emph{impacted users} whose visits are impacted by POI popularity. 
We report the ratio of impacted users averaged over 100 runs of the procedure above (with random selection for dataset splitting). 
As Fig.~\ref{fig:analysis_pop} shows, all datasets have more than 70\% impacted users, which demonstrates the importance   
of POI popularity.

\textbf{Joint impact of co-occurring POIs and POI popularity.} 
The empirical study above confirms the impact of co-occurring POIs and the impact of POI popularity. 
A side observation when comparing Fig.~\ref{fig:analysis_contextual} and Fig.~\ref{fig:analysis_pop} is that these factors 
have a joint impact rather than independent one. In general, for the cities where 
co-occurring POIs  have a greater impact, POI popularity has a less impact (e.g., Edinburgh), and vice versa (e.g., Osaka). 
This brings a challenge 
on designing a model that can learn the impact of the factors jointly and can adapt  
to the different levels of joint impact across different datasets. 

\renewcommand{\arraystretch}{1.0}
\begin{table}
\centering
\small
\caption{Frequently Used Symbols}
\vspace{-3mm}
\label{tab:symbols}
\begin{threeparttable}
\begin{tabular}{c l} 
\toprule[1pt]
Symbol & Description\\ \midrule[1pt]
$\mathcal{R}$ & a set of check-in records\\ \midrule
$\mathcal{L}$ & a set of POIs\\ \midrule
$\mathcal{U}$ & a set of users\\ \midrule
$l$ & a POI\\ \midrule
$u$ & a user\\ \midrule
$s^u$ & a trip of user $u$\\ \midrule
$\vec{l}$ & the latent vector of $l$\\ \midrule
$\vec{u}$ & the latent vector of $u$\\ \midrule
$\vec{c(l)}$ & the latent vector of the co-occurring POIs of $l$\\ \bottomrule[1pt]
\end{tabular}
\end{threeparttable}
\vspace{-4mm}
\end{table}

\section{Problem formulation}
\label{sec:problem}

We aim to learn a context-aware POI embedding such that  POIs co-occurring more frequently are closer in the embedded space. 
We map POIs and users to this embedded space and make trip recommendations based on their closeness in the embedded space.

To learn such an embedding, we use a POI check-in dataset $\mathcal{R}$ (e.g., the datasets summarized in Table~\ref{tab:datasets}). Each check-in record $r\in\mathcal{R}$ is a 3-tuple $\langle u,l,t\rangle$, where $u$ denotes the check-in user, $l$ denotes the POI, and $t$ denotes the check-in time. An example check-in record is $\langle \texttt{10012675@N05},\  \texttt{Art Gallery of Ontario}, \texttt{1142731848}\rangle$, which denotes that user \texttt{10007579@N00} checked-in at \texttt{Art Gallery of Ontario} on \texttt{19 Mar of 2006} (\texttt{1142731848} in UNIX timestamp format).

\textbf{POI visit and historical trip.} 
Let $\mathcal{U}$ be the set of all users and $\mathcal{L}$ be the set of all POIs in the check-in records in $\mathcal{R}$. 
We aggregate a user $u$'s consecutive check-ins at the same POI $l$ into a \emph{POI visit} 
$v^u = \langle u, l, t_a, t_d\rangle$, where $t_a$ and $t_d$ represent the times of the first and the last (consecutive) check-ins at $l$ by $u$.   
With a slight abuse of terminology, we use a POI visit $v^u$ and the corresponding POI $l$ interchangeably as long as the context is clear.  
POI visits of user $u$ within a certain time period (e.g., a day) form a \emph{historical trip} of $u$, denoted as $s^u =\langle v_{1}^u, v_{2}^u, \ldots, v_{|s^u|}^u \rangle$. 
All historical trips of user $u$ form the \emph{profile} of $u$, denoted as $\mathcal{S}^u=\{s_1^u, s_2^u,\dots, s_{|\mathcal{S}^u|}^u\}$.
We learn the POI embedding from the set $\mathcal{S}$ of all historical trips of all users in $\mathcal{U}$, i.e., 
$\mathcal{S} = \mathcal{S}^{u_1} \cup  \mathcal{S}^{u_2} \cup \ldots \cup  \mathcal{S}^{u_{|\mathcal{U}|}}$. 
We summarize the notation in Table~\ref{tab:symbols}.

\textbf{TripRec query.} To showcase the effectiveness of our POI embedding, 
we apply it to a trip recommendation problem~\cite{lim2015personalized}. 
This problem aims to recommend  a \emph{trip} $tr$ formed by an ordered sequence of POIs to a user $u_q$, i.e., $tr = \langle l_1, l_2, \ldots, l_{|tr|} \rangle$, 
such that the value of a \emph{user satisfaction function} is maximized. 
We propose a novel user satisfaction function denoted by $S(u_q,tr)$ which is detailed in Section~\ref{sec:algorithms}.
Intuitively, each POI makes a contribution to $S(u_q,tr)$, and the contribution is larger when the POI suits $u_q$'s preference better. 

A time budget $t_q$ is used to cap the number of POIs in $tr$. The \emph{time cost} of $tr$, denoted by $tc(tr)$, must not 
exceed $t_q$. The time cost $tc(tr)$ 
is the sum of the \emph{visiting time} at every POI $l_i \in tr$, denoted as $tc_v(l_i)$, and  the \emph{transit 
time} between every two consecutive POIs $l_i, l_{i+1} \in tr$, denoted as $tc_t(l_i, l_{i+1})$:
\vspace{-1mm}
\begin{equation}
tc(tr) = \sum_{i=1}^{|tr|}tc_v(l_i) + \sum_{i=1}^{|tr|-1}tc_t(l_i, l_{i+1})
\end{equation}
We derive the visiting time $tc_v(l_i)$ as the average time of POI visits at $l_i$: 

\vspace{-4mm}
\begin{equation}
tc_v(l_i) = \frac{1}{N_{l_i}}\sum_{u\in \mathcal{U}}\sum_{s^u\in \mathcal{S}^u}\sum_{c^u_{j}\in s^u} (v^u_{j}.t_d-v^u_{j}.t_a)\delta(v^u_{j}.l, l_i)
\end{equation}
Here, $N_{l_i}$ represents the total number of POI visits at $l_i$; and $\delta(v^u_{j}.l, l_i)$ is an indicator function that returns 1 if $v^u_{j}.l$ and $l_i$ are the same POI, and 0 otherwise.  
The transit time  $tc_t(l_i, l_{i+1})$ depends on the transportation mode (e.g., by walk or car), which is orthogonal to our study. 
Without loss of generality and following previous studies~\cite{gionis2014customized,lim2015personalized,wang2016improving}, we assume transit by walk 
and derive $tc_t(l_i, l_{i+1})$ as the road network shortest path distance between $l_i$ and $l_{i+1}$ divided by  an average walking speed of 4 km/h. 
Other transit time models can also be used. 

Following~\cite{chen2016learning,lim2015personalized,wang2016improving}, 
we also require $l_1$ and $l_{|tr|}$ to be at a given starting POI $l_s$ and a given ending POI $l_e$. 
For ease of discussion, we call such a trip recommendation problem the \emph{TripRec query}: 

\vspace{-2mm}
\begin{definition}[TripRec Query]
A TripRec query $q$ is represented by a 4-tuple $q=\langle u_q, t_q, l_s, l_e\rangle$.
Given a query user  $u_q$, a query time budget $t_q$, a starting POI $l_s$, and an ending POI $l_e$,  
the TripRec query  finds a trip $tr = \langle l_1, l_2, ..., l_{|tr|}\rangle$ that maximizes $S(u_q,tr)$ and satisfies: (i) $tc(tr) \leqslant t_q$, 
(ii) $l_1 = l_s$, and (iii) $l_{|r|} = l_e$.
\end{definition}

\section{Learning a Context-Aware POI Embedding}
\label{sec:model}

Consider a POI $l_i$, a user $u$, and a historical trip $s$ of $u$ that contains $l_i$.
The popularity of $l_i$, 
the user $u$, and the other  POIs co-occurring in $s$ together form a \emph{context} of $l_i$. Our POI embedding is computed from 
such contexts, and hence is named a \emph{context-aware 
POI embedding}. 
We first discuss how to learn a POI embedding 
such that POIs co-occurring more frequently are closer in the embedded space in Section~\ref{sec:model_contextual}.
We further incorporate user preferences and POI popularities into the embedding in Sections~\ref{sec:model_user} and~\ref{sec:model_pop}. 
We present an algorithm for model parameter learning in Section~\ref{sec:learningalgo}. 

\vspace{-1mm}
\subsection{Learning POI Co-Occurrences}
\label{sec:model_contextual}

Given a POI $l_i$, 
we call another POI $l_j$  a \emph{co-occurring POI} of $l_i$, if 
$l_j$ appears in the same trip as $l_i$. 
The conditional probability $p(l_i|l_j)$, i.e., 
the probability of a trip containing $l_i$
given that $l_j$ is in the trip, 
models the \emph{co-occurrence relationship} of $l_i$ over~$l_j$. 
 
To learn $p(l_i|l_j)$, the Markov model is a solution, which views $p(l_i|l_j)$ as a transition probability from $l_j$ to $l_i$. 
This model assumes that the transition probability of each POI pair is independent from any other POIs, and there are a total of $|\mathcal{L}|^2$ probabilities to be learned. 
Learning such a model requires a large number of check-ins with different adjacent POI combinations. 
This may not be satisfied by real-world POI check-in datasets since check-ins are skewed towards 
popular POIs. Many pairs of POIs may not be observed in consecutive check-ins. 
Learning the transition probability between non-adjacent POIs requires higher-order Markov models which suffers more from data sparsity. 

To overcome the data sparsity problem and capture the co-occurrence relationships between both adjacent and non-adjacent POIs, 
we propose a model to learn $p(l_i|c(l_i))$ instead of $p(l_i|l_j)$, where $c(l_i)$ represents the set of  co-occurring POIs of $l_i$. 
Our model is inspired 
by the \emph{Word2vec} model~\cite{mikolov2013distributed}. 
The Word2vec model embeds words into a vector space where each word is placed in close proximity with its \emph{context words}. Given an occurrence of word $w$ in a large text corpus, each word that 
occurs within a pre-defined distance to $w$ is regarded as a context word of $w$. This pre-defined distance forms a \emph{context window} around a word. 
In our problem, we can view a POI as a ``word'', a historical trip as a ``context window'', the historical trips of a user as a ``document'', and all historical  
trips of all the users as a ``text corpus''. Then, we can learn a POI embedding based on the probability distribution of the co-occurring POIs. 

Specifically, we use the architecture of \emph{continuous bag-of-words} (CBOW)~\cite{mikolov2013efficient}, which predicts the \emph{target word} given its \emph{context}, 
to compute the POI embedding. The computation works as follows. 
Given a POI $l_i\in \mathcal{L}$, we map $l_i$ into a latent $d$-dimensional real space $\mathbb{R}^d$ 
where $d$ is a system parameter,  $d\ll |\mathcal{L}|$. 
The mapped POI, i.e., the  POI embedding, is a $d$-dimensional vector $\vec{l_i}$. 
When computing the embeddings, we treat each historical trip as a context window: given an occurrence of $l_i$ in a historical trip $s$, we treat $l_i$ as the target POI and all other POIs  in $s$ 
as its co-occurring POIs $c(l_i|s)$, i.e., $c(l_i|s) = \{l|l\in s\setminus \{l_i\}\}$.
In the rest of the paper, we abbreviate $c(l_i|s)$ as $c(l_i)$ as long as the context is clear. 

Let $csim(l_i, l_j)$ be the \emph{co-occurrence similarity}  
between two POIs $l_i$ and $l_j$. We compute  $csim(l_i, l_j)$ as 
 the dot product of the embeddings of $l_i$ and $l_j$: 
\begin{equation}
csim(l_i, l_j) = \vec{l_i}\cdot\vec{l_j}
\end{equation}
Similarly, the co-occurrence similarity between a POI $l_i$ and its set of co-occurring POIs $c(l_i)$, denoted as $csim(l_i, c(l_i))$, is computed as: 
\begin{equation}
csim(l_i, c(l_i)) = \vec{l_i}\cdot\vec{c(l_i)}
\end{equation}
Here, $\vec{c(l_i)}$ is computed as an aggregate vector of the embeddings of the POIs in $c(l_i)$. 
We follow Wang et al.~\cite{henry2018vector} and aggregate the embeddings by summing them up in each dimension independently: 
\begin{equation}
\vec{c(l_i)} = \sum_{l\in c(l_i)} \vec{l}
\end{equation}
Other aggregate functions (e.g.,~\cite{wang2015learning}) can also be used.

Then, the probability of observing $l_i$ given $c(l_i)$ is derived by applying the softmax function on the co-occurrence similarity $csim(l_i, c(l_i))$:

\begin{equation}
p(l_i|c(l_i)) = \frac{e^{csim(l_i, c(l_i))}}{Z(\vec{c(l_i)})} = \frac{e^{\vec{l_i}\cdot\vec{c(l_i)}}}{Z(\vec{c(l_i)})} 
\end{equation}
Here, $Z(\vec{c(l_i)}) = \sum_{l\in\mathcal{L}}{e^{\vec{l}\cdot\vec{c(l_i)}}}$ is a normalization term.

\subsection{Incorporating User Preferences}
\label{sec:model_user}

Next, we incorporate user preferences into our model.
We model a user's preferences towards the POIs as her 
 ``co-occurrence'' with the POIs, i.e., a user $u_j$ is also projected 
to a $d$-dimensional embedding space where she is closer to the POIs that she is more likely to visit (i.e., ``co-occur'').
Specifically, the co-occurrence similarity between a POI $l_i$ and a user $u_j$ is computed as:
\begin{equation}
csim(l_i, u_j)=\vec{l_i}\cdot\vec{u_j}
\end{equation}
Thus, the preference of $u_j$ over $l_i$ can be seen as the probability $p(l_i|u_j)$ of observing $l_i$ given $u_j$ in the space. After applying the softmax function over $csim(l_i, u_j)$, $p(l_i|u_j)$ can be computed as:
\begin{equation}
p(l_i|u_j) = \frac{e^{\vec{l_i}\cdot\vec{u_j}}}{Z(\vec{u_j})}
\end{equation}
Here, $Z(\vec{u_j})=\sum_{l\in\mathcal{L}}e^{\vec{l}\cdot\vec{u_j}}$ is a normalization term.

To integrate user preferences with POI co-occurrence relationships, we unify the POI embedding space and the user embedding space into a single embedding space.
In this unified embedding space,  the POI-POI proximity reflects POI co-occurrence relationships and the user-POI proximity reflects user preferences. 
Intuitively, we treat each user  $u_j$ as a ``pseudo-POI''. If user $u_j$ visits POI $l_i$, then $u_j$ (a pseudo-POI) serves as a co-occurring POI of $l_i$. 
Thus, the joint impact of user preferences and POI co-occurrences can be modeled by combining the pseudo POI and the actual co-occurring POIs. 
Given a set of co-occurring POIs $c(l_i)$ and a user $u_j$, the probability of observing $l_i$ can be written as: 
\begin{equation}
p(l_i|c(l_i), u_j)= \frac{e^{\vec{l_i}\cdot(\vec{u_j+c(l_i)})}}{Z(\vec{u_j+c(l_i)})}
\end{equation}
Here, vectors $\vec{u_j}$ and  $\vec{c(l_i)}$ are summed up in each dimension, while 
 $Z(\vec{u_j+c(l_i)}) = \sum_{l\in\mathcal{L}}e^{\vec{l}\cdot(\vec{u_j+c(l_i)})}$ is a normalization term.

\subsection{Incorporating POI Popularity}
\label{sec:model_pop}
We further derive $p(l_i)$ which represents the popularity of $l_i$.
A straightforward model is to count the number of POI visits at $l_i$ and use the normalized frequency as $p(l_i)$.
This straightforward model is used by most existing studies (e.g., ~\cite{gionis2014customized,lim2015personalized,liu2011personalized}). 
This model relies on a strong assumption that POI popularity is linearly proportional to the number of POI visits. 
This linearity assumption may not hold since popularity may not be the only reason for visiting a POI.

Instead of counting POI visit frequency, we propose to learn the POI popularity jointly with 
the impact of co-occurring POIs and user preferences. 
Specifically, we add a dimension to the unified POI and user embedding space, i.e.,  
we embed the POIs to an $\mathbb{R}^{d+1}$ space. This extra dimension represents the latent popularity of a POI, 
and the embedding learned for this space is our \emph{context-aware POI embedding}.

For a POI $l_i$, its embedding now becomes  $\vec{l_i}\oplus l_i.p$ where $\oplus$ is a concatenation operator and $l_i.p$ 
is the latent popularity. The probability $p(l_i)$ is computed by applying the softmax function over $l_i.p$:
\vspace{-1mm}
\begin{equation}
p(l_i) = \frac{e^{l_i.p}}{\sum_{l\in \mathcal{L}}e^{l.p}}
\vspace{-1mm}
\end{equation}
Integrating with the POI contextual relationships and user preferences, the final probability of oberving $l_i$ given $u_j$ and $c(l_i)$ can be represented as:
\vspace{-1mm}
\begin{equation}
p(l_i|c(l_i), u_j)= \frac{e^{\vec{l_i}\cdot(\vec{u_j+c(l_i)})+l_i.p}}{Z(\vec{u_j+c(l_i)}+l_i.p)}
\vspace{-1mm}
\end{equation}
Here, $Z(\vec{u_j+c(l_i)}+l_i.p)=\sum_{l\in\mathcal{L}}e^{\vec{l}\cdot(\vec{u_j+c(l_i)})+l.p}$
is a normalization term.



\subsection{Parameter Learning}
\label{sec:learningalgo}
We adopt the \emph{Bayesian Pairwise Ranking} (BPR) approach~\cite{rendle2009bpr} to learn the embeddings of POIs and users. The learning process aims to maximize the posterior of the observations:
\vspace{-1mm}
\begin{equation}
\Theta = \underset{\Theta}{\mathsf{argmax}}\underset{u\in\mathcal{U}}{\prod}\underset{s^u\in\mathcal{H}^u}{\prod}\underset{l\in s^u}{\prod}\underset{l'\notin s^u}{\prod} P(>_{u,c(l)}|\Theta)p(\Theta)
\vspace{-1mm}
\end{equation}
Here, $\Theta$ represents the system parameters to be learned (i.e., user and POI vectors) and $P(>_{u,c(l)}|\Theta)$ represents the pairwise margin given $u$ and $c(l)$ between the probabilities of observing $l$ and observing $l'$. 
Maximizing the above objective function equals to maximizing its log-likelihood function. Thus, the above equation can be rewritten as follows: 
\vspace{-1mm}
\begin{multline}
\Theta  = \underset{\Theta }{\mathsf{argmax}}\underset{u\in\mathcal{U}}{\sum}\underset{s^u\in\mathcal{H}^u}{\sum}\underset{l\in s^u}{\sum}\underset{l'\notin s^u}{\sum} \log\sigma \Big(\vec{l}\cdot\vec{c(l)}+\vec{l}\cdot\vec{u}+l.p \\
     -\vec{l'}\cdot\vec{c(l)}-\vec{l'}\cdot\vec{u}-l'.p\Big)
\vspace{-1mm}
\end{multline}
Here, $\sigma(\cdot)$ is the sigmoid function and $\sigma(z) = \frac{1}{1+e^{-z}}$.

To avoid overfitting, we add a regularization term $\lambda||\Theta||^2$ to the objective function: 
\begin{multline}
\vspace{-1mm}
\Theta = \underset{\Theta}{\mathsf{argmax}}\underset{u\in\mathcal{U}}{\sum}\underset{s^u\in\mathcal{H}^u}{\sum}\underset{l\in s^u}{\sum}\underset{l'\notin s^u}{\sum} \log\sigma \Big(\vec{l}\cdot\vec{c(l)}+\vec{l}\cdot\vec{u}+l.p \\
    -\vec{l'}\cdot\vec{c(l)}-\vec{l'}\cdot\vec{u}-l'.p\Big) -\lambda||\Theta||^2
\vspace{-1mm}
\end{multline}
%

We use stochastic gradient descent (SGD) to solve the optimization problem. Given a trip $s^u$ of user $u$, we obtain $|s^u|$ observations 
in the form of $\langle u, s^u, l, c(l)\rangle$, where $l \in s^u$. For each observation, we randomly sample $k$ negative POIs not in $s^u$. 
Using each sampled POI $l'$, we update $\Theta$ along the ascending gradient direction:
\vspace{-1mm}
\begin{equation}
\Theta \leftarrow \Theta + \eta \frac{\partial}{\partial\Theta}(\log\sigma(z)-\lambda||\Theta||^2)
\vspace{-1mm}
\end{equation}
Here, $\eta$ represents the learning rate and $z = \vec{l}\cdot\vec{c(l)} + \vec{l}\cdot\vec{u} +l.p - \vec{l'}\cdot\vec{c(l)} - \vec{l'}\cdot\vec{u} - l'.p$ represents the distance between the observed POI and a sampled non-visited POI $l'$.
We summarize the learning algorithm in Algorithm~\ref{alg:learning}, where $itr_m$ represents a pre-defined maximum number of 
learning iterations. 
\begin{algorithm}
\caption{Embedding learning\label{alg:learning}}
\SetKwInOut{Input}{input}
\SetKwInOut{Output}{output}
\SetKwFunction{KwAppend}{append}
\LinesNumbered
\Input{$\mathcal{S}$: a set  of trips; $itr_m$: max iterations}
\Output{$\Theta$}
 Initialize $\Theta$ with Uniform distribution $U(0,1)$\;
 $itr\leftarrow 0$\;
\While{$itr \le itr_m$}{
    \ForEach{observation $\langle u, s^u, l, c(l)\rangle$}{    	
    	Sample a set $\mathcal{L}'$ of $k$ POIs not in $s^u$\;
    	\ForEach{$l'\in \mathcal{L}'$}{
    		$\vec{c(l)}\leftarrow \mathsf{aggregate}(\vec{l_i}|l_i\in c(l))$\;
			$\delta = 1 - \sigma(z)$\;
			$\vec{u}\leftarrow\vec{u} + \eta(\delta(\vec{l}-\vec{l'})-2\lambda\vec{u})$\;  
			$\vec{l}\leftarrow\vec{l} + \eta(\delta(\vec{u}+\vec{c(l)})-2\lambda\vec{l})$\;
			$\vec{l'}\leftarrow\vec{l'} - \eta(\delta(\vec{u}+\vec{c(l)})-2\lambda\vec{l'})$\;
			$l.p\leftarrow \l.p + \eta(\delta-2\lambda l.p)$\;
			$l'.p\leftarrow \l'.p - \eta(\delta-2\lambda l'.p)$\;
			\ForEach{$l_i\in c(l)$}{
				$\vec{l_i}\leftarrow\vec{l_i} + \eta(\delta(\vec{l}-\vec{l'})-2\lambda\vec{l_i})$\;
			}
		}		    		
    }
    $itr \leftarrow itr+1$\;
}
\Return{$\Theta$}\;
\end{algorithm}
\vspace{-5mm}

\section{Trip Recommendation}
\label{sec:algorithms}
To showcase the capability of our context-aware POI embedding to capture the latent POI features, we apply it to 
the TripRec query as defined in Section~\ref{sec:problem}. 

Given a TripRec query $q= \langle u_q, l_s, l_e, t_q\rangle$, the aim is to return a trip $tr = \langle l_1, l_2, \ldots, l_{|tr|}\rangle$ such that 
(i) $tr$ satisfies the query constraints, i.e., starting at $l_s$, ending at $l_e$, and the time cost not exceeding $t_q$ (i.e., $tc(tr) \le t_q$), 
and (ii) $tr$ is most preferred by user $u_q$. 

There may be multiple \emph{feasible trips} that satisfy the query constraints. Let $\mathcal{T}$ be the set of feasible trips. 
The problem then becomes selecting the trip $tr \in \mathcal{T}$ that is  most preferred.
The strategy that guides trip selection plays a critical role in recommendation quality. 

\textbf{Context-aware trip quality score.}
We propose the \emph{context-aware trip quality} (CTQ) score to guide trip selection. 
We thus reduce TripRec to an optimization problem of finding the feasible trip with the highest CTQ score. 
The CTQ score of a trip $tr$, denoted as $S(u_q,tr)$, is a joint score of two factors: the closeness between $tr$ and query $q$ and the co-occurrence similarity among the POIs in $tr$. 
To compute the closeness between $tr$ and $q$, we derive the latent representation $\vec{q}$ of $q$ as an aggregation (e.g., summation) of the vectors $\vec{u_q}$, $\vec{l_s}$, and $\vec{l_e}$. 
The closeness between $q$ and a POI $l$, denoted as $clo(q,l)$, is computed as the probability of observing $l$ given $q$:
\vspace{-2mm}
\begin{equation}
clo(q,l) = \frac{e^{\vec{l}\cdot\vec{q}}}{\sum_{l'\in\mathcal{L}}e^{\vec{l'}\cdot\vec{q}}}
\vspace{-1mm}
\end{equation}
The closeness between $q$ and $tr$, denoted as $clo(q,tr)$, is the sum of $clo(q,l)$ for every $l \in tr$:
\vspace{-2mm}
\begin{equation}
clo(q,tr) = \sum_{i=2}^{|tr|-1} clo(q, l_i)
\vspace{-1mm}
\end{equation}

The co-occurrence similarity among the POIs in $tr$ is computed as the sum of the pairwise \emph{normalized occurrence similarity} between any two POIs $l_i$ and $l_j$ in $tr$,
denoted as $ncsim(l_i, l_j)$: 
\vspace{-1mm}
\begin{equation}
ncsim(l_i, l_j)=e^{\vec{l_i}\cdot\vec{l_j}}/\sum_{l}\sum_{l':l'\neq l} e^{\vec{l}\cdot\vec{l'}}
\vspace{-1mm}
\end{equation}
Overall, the CTQ score $S(u_q, tr)$ is computed as:
\vspace{-1mm}
\begin{equation}
S(u_q,tr)=\sum_{i=2}^{|tr|-1} \frac{e^{\vec{q}\cdot\vec{l_i}}}{\sum_{l} e^{\vec{q}\cdot\vec{l}}}+ \sum_{i=2}^{|tr|-2}\sum_{j=i+1}^{|tr|-1} \frac{e^{\vec{l_i}\cdot\vec{l_j}}}{\sum_{l}\sum_{l':l'\neq l} e^{\vec{l}\cdot\vec{l'}}}
\end{equation}
Here, we have omitted $l_1$ and $l_{|tr|}$. This is because all feasible trips share the same $l_1$ and $l_{|tr|}$ which are the given starting and ending POIs $l_s$ and $l_e$ in the query.

\textbf{Problem reduction.}
To generate the feasible trips, we construct a directed graph $G=(V,E)$, where each vertex  $v_i \in V$ represents POI $l_i \in \mathcal{L}$ and each edge $\overrightarrow{e_{ij}} \in E$ represents the 
transit from $v_i$ to $v_j$. We assign \emph{profits} to the vertices and edges. The profit of vertex $v_i$, denoted as $f(v_i)$, is computed as $f(v_i)=clo(q,l_i)$. 
The profit of an edge $\overrightarrow{e_{ij}}$, denoted as $f(\overrightarrow{e_{ij}})$, is computed as $f(\overrightarrow{e_{ij}} )=ncsim(l_i,l_j)$. 
For ease of discussion, we use $v_1$ and $v_{|V|}$ to represent the query starting and ending POIs $l_s$ and $l_e$, respectively. 
We set the profits of $v_1$ and $v_{|V|}$ as zero, since they are included in every feasible trip. 
We further add \emph{costs} to the edges to represent the trip cost. The cost of edge $\overrightarrow{e_{ij}} $, denoted as $tc(\overrightarrow{e_{ij}})$, is the sum of the transit time cost between $l_i$ and $l_j$ 
and the visiting time cost of $l_j$, i.e.,  $tc(\overrightarrow{e_{ij}} )=tc_v(l_j)+tc_t(l_i,l_j)$. 

Based on the formulation above, recommending a trip for query $q$ can be seen as a variant of the \emph{orienteering problem}~\cite{golden1987orienteering} which finds a path that collects the most profits in $G$ while costs no more than a given budget $t_q$. We thus reduce the  TripRec problem to the following constrained optimization problem:
\begin{equation}\label{eq:problem}
\begin{array}{l}
\displaystyle \text{max}\  \sum_{i=1}^{|V|}\sum_{j=1}^{|V|} x_{ij}\cdot f(v_j)+\sum_{i=2}^{|V|-1}\sum_{j=i+1}^{|V|-1}x_i\cdot x_j\cdot f(\overrightarrow{e_{ij}}) \\
\text{s.t.}\ \   \displaystyle \text{(a) } \sum_{i=1}^{|V|} x_{1i} = x_1=1, \quad  \text{(b) }  \sum_{i=1}^{|V|} x_{i|V|} = x_{|V|}=1\\
 \quad  \quad      \displaystyle     \text{(c) }    \sum_{j=1}^{|V|} x_{ij} = \sum_{k = 1}^{|V|} x_{ki} = x_i \leqslant 1,\ \forall i\in [2,|V|-1]\\
\quad  \quad      \displaystyle    \text{(d) }       tc_v(v_1) + \sum_{i=1}^{|V|-1}\sum_{j=2}^{|V|} x_{ij}\cdot tc(\overrightarrow{e_{ij}}) \leqslant t_q\\          
 \quad \quad     \displaystyle      \text{(e) }      2\leqslant p_i\leqslant |V|,\ \forall i\in [2,|V|]\\
 \quad \quad       \displaystyle   \text{(f) }       p_i - p_j +1\leqslant (|V|-1)(1-x_{ij}),\ \forall i,j\in [2,|V|]
\end{array}
\end{equation}
Here, $x_{ij}$ and $x_i$ are boolean indicators: $x_{ij}=1$ if edge $\overrightarrow{e_{ij}}$ is selected, and $x_i= 1$ if vertex $v_i$ is selected. 
Conditions (a) and (b) restrict the trip to start from $v_1$ and end at $v_{|V|}$. Condition (c) restricts to visit any selected POI once. 
Condition~(d) denotes the time budget constraint. Conditions~(e) and~(f) are adapted from~\cite{miller1960integer}, where $p_i$ denotes the position of $v_i$ in the trip. They ensure no cycles in the trip.

\subsection{The C-ILP Algorithm}\label{sec:cilp}
\vspace{-1mm}
A common approach for the orienteering problems is the \emph{integer linear programming} (ILP) algorithm~\cite{chen2016learning,lim2015personalized}. 
However, ILP does not apply directly to our problem. This is because the second term in our objective function in Equation~\ref{eq:problem}, i.e., 
$\sum_{i=2}^{|V|-1}\sum_{j=i+1}^{|V|-1}x_i\cdot x_j\cdot f(\overrightarrow{e_{ij}})$, is nonlinear.
In what follows, we transform Equation~\ref{eq:problem} to a linear form such that the ILP algorithm~\cite{chen2016learning,lim2015personalized} 
can be applied to solve our problem. Such an algorithm finds 
the exact optimal trip for TripRec. We denote it as the \emph{C-ILP} algorithm for ease of discussion. 

Our transformation replaces the vertex indicators $x_i$ and $x_j$ in Equation~\ref{eq:problem}  
with a new indicator $x'_{ij}$, where  $x'_{ij} = 1$ if both $v_i$ and $v_j$ are selected (not necessarily adjacent). 
We further impose  $i<j$ in $x'_{ij}$ to reduce the total number of such indicators by half. This does not affect the correctness of the optimization 
since $x'_{ij} = x'_{ji}$.
Then, Equation~\ref{eq:problem} is rewritten as follows.
\vspace{-1.5mm}
\begin{equation}\label{eq:problem2}
\begin{array}{l}
\text{max}\ \sum_{i=1}^{|V|}\sum_{j=1}^{|V|} x_{ij}\cdot f(v_j) + \sum_{i=2}^{|V|-2}\sum_{j=i+1}^{|V|-1}x'_{ij}\cdot f(\overrightarrow{e_{ij}})\\
\text{s.t.} \ \   \displaystyle \text{(a) } \sum_{i=1}^{|V|} x_{1i} = 1, \quad     \text{(b) } \sum_{i=1}^{|V|} x_{i|V|} = 1\\
\quad \quad         \displaystyle     \text{(c) } \sum_{j=1}^{|V|} x_{ij} = \sum_{k = 1}^{|V|} x_{ki} \leqslant 1,\ \forall i\in [2,|V|-1]\\
\quad \quad         \displaystyle     \text{(d) } x'_{ij} = \sum_{k=1}^{|V|}\sum_{m=1}^{|V|} x_{ik}\cdot x_{jm},\ \forall i,j\in [1, |V|-1],i<j\\
\quad \quad         \displaystyle     \text{(e) } x'_{i|V|} = \sum_{k=1}^{|V|} x_{ik}, \forall i\in [1,|V|-1]\\
\quad \quad         \displaystyle     \text{(f) } tc_v(v_1) + \sum_{i=1}^{|V|}\sum_{j=1}^{|V|} x_{ij}\cdot tc(\overrightarrow{e_{ij}}) \leqslant t_q\\
\quad \quad         \displaystyle     \text{(g) }2\leqslant p_i\leqslant |V|,\ \forall i\in [2,|V|]\\
\quad \quad         \displaystyle     \text{(h) } p_i - p_j +1\leqslant (|V|-1)(1-x_{ij}), \ \forall i,j\in [2,|V|]
\end{array}
\vspace{-1.5mm}
\end{equation}
Here, Conditions (a) to (c) and (f) to (h) are the same as those in Equation~\ref{eq:problem}. Conditions (d) and (e) define the relationships between $x_{ij}$ and $x'_{ij}$. The main idea is that if a trip includes a vertex $v_i$, it must contain an edge starting from $v_i$, or an edge ending at $v_i$ if $v_i=v_{|V|}$.
Thus, for any two vertices $v_i$ and $v_j$ that are not $v_{|V|}$, there indicator $x_{ij}$ equals to 1 if the solution trip contain two edges:one starting from $v_i$ and another from $v_j$. For any vertex $v_i$ and the vertix $v_{|V|}$, their indicator $x_{ij}$ equals to 1 if the solution trip contains an edge starting from $v_i$.
Using $x'_{ij}$, we transform our  objective function into a linear form. 
 Condition (d) is still non-linear (note $x_{ik}\cdot x_{jm}$). We replace it with three linear constraints:
\vspace{-2mm}
\begin{equation}
\label{eq:transformation}
\begin{array}{l}
  \displaystyle  x'_{ij}\leqslant \sum_{k=1}^{|V|}x_{ik},\forall i,j\in [1,|V|-1], i<j\\
  \displaystyle x'_{ij}\leqslant \sum_{k=1}^{|V|}x_{jk},\forall i,j\in [1,|V|-1], i<j\\
  \displaystyle x'_{ij}\geqslant \sum_{k=1}^{|V|}\sum_{m=1}^{|V|} (x_{ik} + x_{jm}) -1,  \forall i,j\in [1,|V|-1], i<j
 \end{array}
\end{equation}
To show the correctness of the above transformation, 
we consider two cases: (i) At least one vertex (e.g., $v_i$) is not included in the optimal trip; and (ii) Both  $v_i$ and $v_j$ are not in the optimal trip. Condition (d) in Equation~\ref{eq:problem2} ensures that $x'_{ij} = 0$ in Case (i) and $x'_{ij}=1$ in Case (ii). We next show that this is also guaranteed by Equation~\ref{eq:transformation}. For Case (i), we have $\sum_{k=1}^{|V|}x_{ik}=0$, which leads to $x'_{ij} \leqslant 0$ according to the first constraint in Equation~\ref{eq:transformation}. Since $x'_{ij}\in [0,1]$, we have $x'_{ij}=0$. For Case (ii), we have $\sum_{k=1}^{|V|}x_{ik}=1$ and $\sum_{k=1}^{|V|}x_{jk}=1$. According to the third constraint in Equation~\ref{eq:transformation}, we have $x'_{ij}\geqslant 1$. Since $x'_{ij}\in [0,1]$, we have $x'_{ij}=1$. Combining the two cases, we show that the above transformation retains the constraints
of Condition (d).  

\textbf{Algorithm complexity.}
There are $2\cdot |E|$ boolean variables in C-ILP, where $|E|$ represents the number of edges in $G$. To compute the solution, the \emph{lpsolve} algorithm~\cite{berkelaar2004lpsolve}  first finds a trip without considering the integer constraints, which can be done in $O(|E|)$ time. Then it refines the trip to find the optimal integral solution. Given a non-integer variable in the current trip, the algorithm splits the solution space into two: one restricting the variable to have at least the ceiling of its current value and the other restricting the variable to have at most the floor of its current value. Then, the algorithm optimizes the two solution spaces and checks if there still exist non-integer variables in the new trip. The algorithm repeats the above procedure until an integral solution is found.   The algorithm uses branch-and-bound to guide the search  process. It may need to explore all possible combinations in the worst case, which leads to a worst-case time complexity of $O(2^{|E|})$.

\subsection{The C-ALNS Algorithm}\label{sec:calns}
The C-ILP algorithm finds the trip with the highest CTQ score. However, the underlying 
integer linear program algorithm may incur a non-trivial running time as shown by the complexity analysis above.

To avoid the high running time of C-ILP, 
we propose a heuristic algorithm named C-ALNS that is based on \emph{adaptive large neighborhood search} (ALNS)~\cite{pisinger2007general}. 
 ALNS  is a meta-algorithm to generate heuristic solutions. It starts with an initial solution (a trip in our problem) and then improves the solution iteratively 
by applying a destroy and a build operator in each iteration. 
The \emph{destroy operator} randomly removes a subset of the elements (POIs) from the current solution. 
The \emph{build operator} inserts new elements  into the solution to form a new solution. Different destroy/build operators use different heuristic strategies to select the elements to remove/insert.
Executing a pair of destroy and build operators can be viewed as a move to explore a  neighborhood of the current solution. 
The aim of the exploration is to find a solution with a higher objective function value. The algorithm terminates after a pre-defined maximum number of 
iterations $itr_m$ is reached.

As summarized in Algorithm~\ref{alg:ALNS}, 
our C-ALNS algorithm adapts the ALNS framework as follows:
(i) C-ALNS consists of multiple ($run_m$) ALNS runs (Lines~\ref{algl:ALNS run start}). The best trip of all runs and its CTQ score are stored as $tr_{opt}$ and $S(u_q, tr_{opt})$. 
The best trip within a single run is stored as $tr_{r\_opt}$. The algorithm initializes a solution pool $\mathcal{P}$ (detailed in Section~\ref{sec:solutionIni}) before running ALNS, 
where a trip  from the solution pool is randomly selected to serve as the initial solution of ALNS. 
(ii) C-ALNS uses multiple pairs of destroy operators $\mathcal{D}$ and build operators $\mathcal{B}$ to enable random selections of the operators to be used ALNS (detailed in Section~\ref{sec:operators}).
(iii) C-ALNS uses a  local search procedure after a new solution is built to explore different visiting orders over the same set of POIs (detailed in Section~\ref{sec:localSearch}).
(iv)~C-ALNS uses a \emph{Simulated Annealing} (SA) strategy  to avoid falling in local optimum (detailed in Section~\ref{sec:SA}).

\vspace{-2mm}
\begin{algorithm}
\caption{C-ALNS\label{alg:ALNS}}
\SetKwInOut{Input}{input}
\SetKwInOut{Output}{output}
\SetKwFunction{KwAppend}{append}
\LinesNumbered
\Input{POI Graph $G$, Query $q=\langle u_q, l_s, l_e, t_q\rangle$}
\Output{Optimal trip $tr_{opt}$}
$tr_{opt}\leftarrow \emptyset, S(u_q, tr_{opt})\leftarrow -\infty, run \leftarrow 0$\;
initialize the solution pool $\mathcal{P}$\;
\While{$ run \le run_m$\label{algl:ALNS run start}}{
	$tr\leftarrow \mathsf{RandomSelect}(\mathcal{P})$\label{algl:solutionIni}\;
	$tr_{r\_opt}\leftarrow tr$\;
	$temp\leftarrow \tau$\label{algl:tempIni}\;	
	initialize the weights $\mathcal{D}$ and $\mathcal{B}$\;
	$itr \leftarrow 0$\;
    \While{$itr \le itr_m$}{  
    	
    	$\{d,b\}\leftarrow \mathsf{RandSelect}(\mathcal{D},\mathcal{B})$\;    	
    	$tr'\leftarrow\mathsf{Apply}(tr, d)$\;	    	
    	$tr'\leftarrow\mathsf{Apply}(tr, b)$\;
    	$\mathsf{LocalSearch}(tr')$\label{algl:localsearch}\;
    	\If{$S(u_q, tr')>S(u_q, tr)\ \text{or} \ x^{U(0,1)}<exp(\frac{S(u_q, tr')-S(u_q, tr))}{temp})$\label{algl:solution_accept}} {
			$tr\leftarrow tr'$\;  
			\If{$S(u_q, tr_{r\_opt})<S(u_q, tr)$}{
				$tr_{r\_opt}\leftarrow tr$\;			
			}  	
    	}
    	$temp\leftarrow temp\times \theta$\label{algl:cooling}\;
    	update the weights of $\mathcal{D}$ and $\mathcal{B}$\; 		
    }
    \If{$S(u_q, tr_{opt})<S(u_q, tr_{r\_opt})$}{
		$tr_{opt}\leftarrow tr_{r\_opt}$\;
	}
	update $\mathcal{P}$\;
}\label{algl:ALNS run end}
\Return{$tr_{opt}$}
\end{algorithm}

\vspace{-2mm}
\subsubsection{The Solution Pool\label{sec:solutionIni}} 
We maintain a subset of feasible trips in the solution pool $\mathcal{P}$, where each trip $tr_i$ is stored with its CTQ score as a tuple: $\langle tr_i, S(u_q, tr_i)\rangle$.
At the beginning of each ALNS run, we select a trip from the solution pool $\mathcal{P}$ and use it as the initial trip for the run. The probability of selecting a trip $tr_i$ is computed as $p(tr_i)=S(u_q, tr_i)/\sum_{tr \in \mathcal{P}} S(u_q, tr)$. 
At the end of each run, we insert the tuple $\langle tr_{r\_opt}, S(u_q, tr_{r\_opt})\rangle$ into $\mathcal{P}$, where $tr_{r\_opt}$ is the best trip accepted in this run. 
We keep $N$ trips with the highest CTQ scores in $\mathcal{P}$, where $N$ is a system parameter.

We initialize the solution pool with three initial trips generated by a low-cost heuristic based algorithm. 
This algorithm  first creates a trip with  two vertices $v_1$ and $v_{|V|}$ corresponding to the starting and ending POIs $l_s$ and $l_e$ of the query. 
Then, it iteratively inserts a new vertex into the trip until the time budge is reached. 
To choose the next vertex to be added, we use the following three different strategies, yielding the three initial trips: 
\begin{itemize}
\item
Choose the vertex $v$ that adds the highest profit to maximize $f^+_{\Delta}(v) = S(u_q, tr')-S(u_q,tr)$, where $tr$ is the current trip and $tr'$ is the trip after adding~$v$.
\item
Choose the vertex $v$ that adds the least time cost to minimize $tc^+_{\Delta}(v)$, where $t^+_{\Delta}(v) = tc(tr')-tc(tr)$. 
\item
Choose the most cost-effective vertex $v$ that maximizes $f^+_{\Delta}(v)/t^+_{\Delta}(v)$.
\end{itemize}

\subsubsection{The Destroy and Build Operators\label{sec:operators}}  
\textbf{The destroy operator.} Given a trip $tr$ and a removal fraction parameter $\rho\in [0,1]$, a destroy operator removes $\lceil \rho\cdot (|tr|-2)\rceil$ vertices from $tr$. 
We use four destroy operators with different removal strategies:

\emph{Random.} This operator randomly selects $\lceil \rho\cdot (|tr|-2)\rceil$ vertices to be removed.

\emph{Least profit reduction.} This operator selects $\lceil \rho\cdot (|tr|-2)\rceil$ vertices with the least profit reduction: $f^-_{\Delta}(v)=S(u_q, tr)-S(u_q,tr')$, where $tr'$ represents the trip 
after $v$ is removed from $tr$. We add randomness to this operator. Given the list of vertices in $tr$ sorted in ascending order of their profits, we compute the next vertex to be removed as $(x^{U(0,1)})^{\psi\cdot\rho}(|tr|-2)$. Here, $x$ is a random value generated from the Uniform distribution $U(0,1)$ and the parameter $\psi$ is a system parameter that represents the extent of randomness imposed on this operator. A larger value of $\psi$ leads to less randomness.

\emph{Most cost reduction.} This operator selects $\lceil \rho\cdot (|tr|-2)\rceil$ vertices with the largest cost reduction: $t^-_{\Delta}(v)$. We also randomize this operator in the same way as 
the least profit reduction operator. 

\emph{Shaw removal.} This operator implements the Shaw removal~\cite{ropke2006adaptive}. It randomly selects a vertex $v$ in $tr$ and removes $\lceil \rho\cdot (|tr|-2)\rceil$ vertices with the smallest distances to $v$. We also randomize this operator as we do above. 

\textbf{The build operator.}  The build operator adds vertices to $tr$ until the time budget is reached. We use four build operators as follows.

\emph{Most profit increment.} This operator iteratively inserts an unvisited vertex that adds the most profit.

\emph{Least cost increment.} This operator iteratively inserts an unvisited vertex that adds the least time cost.

\emph{Most POI similarity.} This operator randomly selects a vertex $v_i$ in $tr$. Then, it sorts the unvisited vertices by their distances to $v_i$ 
in our POI embedding space. The unvisited vertices nearest to $v_i$ are added to $tr$.

\emph{Highest potential.} This operator iteratively inserts an unvisited vertex $v_i$ that, together with another unvisited vertex $v_j$,  adds the most profit while the two vertices  
 do not exceed the  time budget.

\textbf{Operator choosing.} 
We use a roulette-wheel scheme to select the operators to be applied. Specifically, we associate a weight $w$ to each destroy or build operator, which represents its performance in previous iterations to increase  the CTQ score. The probability of selecting an operator $o_i$ equals to its normalized weight (e.g., $o_i.w/\sum_{o\in\mathcal{D}} o.w$ if $o_i$ is a destroy operator).
At the beginning of each ALNS run, we initialize the weight of each operator to be 1. After each iteration in a run, we score the applied operators based on their performances. 
We consider four scenarios: (i) a new global best trip $tr_{opt}$ is found; (ii) a new local best trip within the run is found; (iii) a local best trip within the run is found but it is not new; and (iv) 
the new trip  is worse than the previous trip but is accepted by the Simulated Annealing scheme. 
We assign different scores for different scenarios. The operator scoring scheme is represented as a vector $\vec{\pi} = \langle \pi_1, \pi_2, \pi_3, \pi_4, \pi_5 \rangle $, where each element corresponds to a scenario, e.g., $\pi_1$ represents the score for Scenario (i), and $\pi_5$ corresponds to any scenario not listed above. We require $\pi_1>\pi_2>\pi_3>\pi_4>\pi_5$. 
Given an operator $o_i$, its current weight $o_i.w$ and its score $o_i.\pi$, we update the weight of $o_i$ as $o_i.w\leftarrow\kappa\cdot o_i.w + (1-\kappa)\cdot o_i.w$. 
Here, $\kappa$ is a system parameter controlling the weight of the scoring action. 

\subsubsection{Local Search\label{sec:localSearch}}
The local search function $\mathsf{LocalSearch}$ (Algorithm~\ref{alg:ALNS}, Line~\ref{algl:localsearch})  
takes a trip $tr$ as its input and explores trips that consist of the same set of vertices of $tr$  
but have different visiting orders. We adapt the \emph{2-opt edge exchange} technique for efficient local exploration. Specifically, the 2-opt edge exchange procedure iteratively performs the following procedure: 
(i) remove two edges from $tr$; (ii) among the three sub-trips produced by Step (i), reverse the visiting order of the second sub-trip; (iii) reconnect the three sub-trips. 
For example, let $tr = \langle v_1, v_2, v_3, v_4, v_5, v_6 \rangle$. Assume that we remove  edges $\overrightarrow{e_{1,2}}$ and $\overrightarrow{e_{4,5}}$, which results in three sub-trips $\langle v_1 \rangle$, 
$\langle v_2, v_3, v_4 \rangle$, and $\langle v_5, v_6\rangle$. We swap the visiting order of the second sub-trip and reconnect it with the other two sub-trips, producing a new trip  
$\langle v_1, v_4, v_3, v_2, v_5, v_6\rangle$. If the new trip has a lower time cost, we accept the change and proceed to next pair of unchecked edges.

\renewcommand{\arraystretch}{1}
\begin{table*}
\vspace{-1mm}
\centering
\scriptsize
\caption{Performance Comparison in Recall, Precision, and F$_1$-score\label{tab:result1}}
\vspace{-3mm}
\begin{tabular}{|l|*3c|*3c|*3c|*3c|}
\hline
City &  \multicolumn{3}{|c|}{Edin.} &  \multicolumn{3}{|c|}{Glas.} & \multicolumn{3}{|c|}{Osak.} & \multicolumn{3}{|c|}{Toro.}  \\ \hhline{|-|---|---|---|---|}
Algorithm & Rec. & Pre. & F$_1$ & Rec. & Pre. & F$_1$ & Rec. & Pre. & F$_1$ & Rec. & Pre. & F$_1$ \\ \hhline{|-|---|---|---|---|}
Random & 0.052 & 0.079 & 0.060 & 0.071 & 0.092 & 0.078 & 0.057 & 0.074 & 0.063 & 0.045 & 0.060 & 0.050 \\ 
Pop & 0.195 & 0.238 & 0.209 & 0.104 & 0.128 & 0.112  & 0.110 & 0.138 & 0.121 & 0.114 & 0.148 & 0.125  \\ 
MF & 0.242 & 0.229 & 0.233 & 0.310 & 0.308 & 0.307 & 0.195 & 0.173 & 0.181& 0.408 & 0.410 & 0.407   \\ 
PersTour & 0.455 & 0.418 & 0.430 & 0.589 & 0.571 & 0.577 & 0.406 & 0.384 & 0.392 & 0.431 & 0.422 & 0.425 \\ 
POIRank & 0.326 & 0.326 & 0.326 & 0.408 & 0.408 & 0.408  & 0.367 & 0.367	 & 0.367 & 0.389	& 0.389 & 0.389\\ 
M-POIRank & 0.318 & 0.318 & 0.318 & 0.387 & 0.387 & 0.387 & 0.328 & 0.328 & 0.328 & 0.379 & 0.379 & 0.379  \\ 
C-ILP (proposed) & \textbf{0.555} & \textbf{0.527} & \textbf{0.538} & \textbf{0.659} & \textbf{0.646} & \textbf{0.651} & \textbf{0.497} & \textbf{0.492} & \textbf{0.494} & \textbf{0.618} & \textbf{0.601} & \textbf{0.608}\\ 
C-ALNS (proposed) & \textit{0.554} &\textit{0.527} &  \textit{0.537} & \textit{0.657} & \textit{0.645} & \textit{0.649} & \textit{0.496} & \textit{0.491} & \textit{0.493} & \textit{0.616} & \textit{0.598} & \textit{0.607} 
\\ \hline
\end{tabular}
\vspace{-2mm}
\end{table*} 

\renewcommand{\arraystretch}{1}
\begin{table*}
\centering
\scriptsize
\caption{Performance Comparison in Recall$^*$, Precision$^*$, and F$_1^*$-score\label{tab:result2}}
\vspace{-3mm}
\begin{tabular}{|l|*3c|*3c|*3c|*3c|}
\hline
City & \multicolumn{3}{|c|}{Edin.} & \multicolumn{3}{|c|}{Glas.} & \multicolumn{3}{|c|}{Osak.}  &  \multicolumn{3}{|c|}{Toro.} \\ \hhline{|-|---|---|---|---|}
Algorithm & Rec$^*$. & Pre$^*$. & F$_1^*$ & Rec$^*$. & Pre$^*$. & F$_1^*$ & Rec$^*$. & Pre$^*$. & F$_1^*$ & Rec$^*$. & Pre$^*$. & F$_1^*$ \\ \hhline{|-|---|---|---|---|}
PersTour & 0.740 & 0.633 & 0.671 & 0.826 & 0.782 & 0.798 & 0.759 & 0.662 & 0.699  & 0.779 & 0.706 & 0.732  \\ 
POIRank & 0.700 & 0.700 & 0.700 & 0.768 & 0.768 & 0.768 & 0.745 & 0.745 & 0.745  & 0.754 & 0.754 & 0.754  \\ 
M-POIRank & 0.697 & 0.697 & 0.697 & 0.762 & 0.762 & 0.762  & 0.732 & 0.732 & 0.732 & 0.751 & 0.751 & 0.751 \\ 
C-ILP (proposed) & \textbf{0.792} & \textbf{0.754} & \textbf{0.769} & \textbf{0.864} & \textbf{0.844} & \textbf{0.853}  & \textbf{0.793} & \textbf{0.740} & \textbf{0.763}  & \textbf{0.842} & \textbf{0.800} & \textbf{0.818}  \\ 
C-ALNS (proposed) & \textit{0.792}& \textit{0.752} & \textit{0.768}   & \textit{0.862} & \textit{0.843} & \textit{0.852}  & \textit{0.792} & \textit{0.739} & \textit{0.762}  & \textit{0.841} & \textit{0.798} & \textit{0.815}
\\ \hline
\end{tabular}
\vspace{-5mm}
\end{table*}

\subsubsection{Simulated Annealing\label{sec:SA}}
We adapt the \emph{simulated annealing} (SA) technique to avoid  local optima. 
Specifically, at the beginning of  each ALNS run, we initialize a temperature $temp$ to a pre-defined value $\tau$.
After every iteration, a new trip $tr'$ is generated from a previous trip $tr$. If $S(u_q, tr') < S(u_q, tr)$, we do not discard $tr'$ immediately. 
Instead, we further test whether $S(u_q, tr') - S(u_q, tr) > -temp\times\log x$ where $x$ is a random value generated from the Uniform distribution $U(0,1)$ (Algorithm~\ref{alg:ALNS}, Line~\ref{algl:solution_accept}). 
If yes, we still replace $tr$ with $tr'$.  We gradually reduce the possibility of keeping a worse new trip by decreasing the value of $temp$ after each iteration by a pre-defined cooling factor $\theta$ (Algorithm~\ref{alg:ALNS}, Line~\ref{algl:cooling}). 

\textbf{Algorithm complexity.} 
 C-ALNS  has $run_m$ ALNS runs, where each run applies $itr_m$ pairs of destroy-build operators. To apply a detroy operator, the algorithm needs to perform $|tr_{avg}|$ comparisons to choose the vertices to remove. To apply a build operator, the algorithm needs to perform $|V|$ comparisons to choose the vertices to add. Here, $|tr_{avg}|$ represents the average length of feasible trips and $|V|$ represents the number of vertices in $G$. Thus, the time complexity of C-ALNS is $O(run_m\cdot itr_m\cdot (|tr_{avg}| + |V|))$.

\section{Experiments}
\label{sec:experiments}
We evaluate the effectiveness and efficiency of the proposed algorithms empirically in this section.
We implement the algorithms in Java. We run the experiments on a 64-bit Windows machine with 24 GB memory and a 3.4 GHz Intel Core i7-4770 CPU.

\subsection{Settings}
We use four real-world POI check-in datasets from Flickr (cf. Section~\ref{sec:empirical}). {\color{red} }
We perform leave-one-out cross-validation on the datasets. 
In particular, we use a  trip of a user $u$ with at least three POIs as a testing trip $tr^*$.  
We use $u$ as the query user,  the starting and ending POIs of $tr^*$ as the query starting and ending POIs, and the time cost of $tr^*$ as the query time budget. 
We use all the other trips in the dataset for training to obtain the context-aware embeddings for the POIs and $u$. 

Let $tr$ be a trip recommended by an algorithm. 
We evaluate the algorithms with three metrics: 
(i) \emph{Recall} -- the percentage of the POIs in $tr^*$
that are also in $tr$, (ii) \emph{Precision} -- the 
percentage of the POIs in $tr$ that are also in $tr^*$, 
(iii) \emph{F$_1$-score} -- the harmonic mean of Precision and Recall. 
We exclude the starting and ending POIs when computing these three metrics. 
To keep consistency with two baseline algorithms~\cite{chen2016learning,lim2015personalized}, 
we further report three metrics denoted as \emph{Recall$^*$}, \emph{Precision$^*$}, and \emph{F$_1^*$-score}. 
These metrics are counterparts of Recall, Precision, and F$_1$-score, but they include the starting and ending POIs in the computation.  

We test both our algorithms \textbf{C-ILP} (Section~\ref{sec:cilp}) and \textbf{C-ALNS} (Section~\ref{sec:calns}).  
They use the same context-aware POI embeddings as described in Section~\ref{sec:model}.
We learn a 13-dimensional embedding with a learning rate $\eta$ of 0.0005 and a regularization term parameter $\lambda$ of 0.02. 
For  C-ALNS, we set the removal fraction $\rho$ as 0.2, the operator scoring vector $\vec{\pi}$ as $\langle 10,5,3,1,0\rangle$, the SA initial temperature as 0.3, and the cooling factor as 0.9995.

\textbf{Baseline algorithms.}
We compare with the following  five baseline algorithms:

\textbf{Random.} This algorithm repeatedly adds a randomly chosen unvisited POI to the recommended trip until reaching the query time budget. 

\textbf{Pop.} This algorithm repeatedly adds the most popular unvisited POI to the recommended trip until reaching the query time budget.  
The popularity of a POI is computed as the normalized POI visit frequency.

\textbf{MF.} This algorithm repeatedly adds the unvisited POI with the highest \emph{user interest score} 
to the recommended trip until reaching the query time budget.  
The user interest score of a POI is computed using \emph{Bayesian Probabilistic Matrix Factorization}~\cite{salakhutdinov2008bayesian} over the matrix of users and POI visits.

 \textbf{PersTour~\cite{lim2015personalized}.} This algorithm recommends the trip that meets the time budget 
and has the highest sum of \emph{POI scores}. 
The POI score of a POI $l$ is the weighted sum of its popularity and user interest score, where the popularity is computed with the same method as in \textbf{Pop.}, and the user interest score is derived from the query user's previous visiting durations at POIs with the same category as $l$. We use a weight of 0.5, which is reported to be optimal~\cite{lim2015personalized}.


\textbf{POIRank~\cite{chen2016learning}.} This algorithm resembles PersTour but differs in how the POI score is computed. 
It represents each POI as a feature vector of five dimensions: POI category, neighborhood, popularity, visit counts, and visit duration (cf.~Section~\ref{sec:lit_gen}). 
It computes the POI score of each POI using \emph{rankSVM} with linear kernel and L2 loss~\cite{lee2014large}. 
We further test its variant \textbf{M-POIRank} where a weighted \emph{transition score} is added to the POI score.
Given a pair of POIs, their transition score is modeled using the Markov model that factorizes the transition probability between the two POIs as 
the product of the transition probabilities between the five POI features of the two POIs.

\subsection{Results}
\emph{Overall performance.}
We summarize the results in Tables~\ref{tab:result1} and~\ref{tab:result2} (Random, Pop, and MF are uncompetitive and are 
omitted in Table~\ref{tab:result2} due to space limit). We highlight the best result in \textbf{bold} and the second best result in \textit{italics}. We see that both C-ILP and C-ALNS consistently outperform the baseline algorithms. 
C-ILP outperforms PersTour, the baseline with the best performance, by $25\%, 13\%, 26\%$, and $43\%$ in F$_1$-score on the datasets Edinburgh, Glasgow, Osaka, and Toronto, respectively. 
 C-ALNS has slightly lower scores than those of C-ILP, but the difference is very small (0.002 on average). 
This confirms  the capability of our heuristic algorithm C-ALNS to generate high quality trips.  

\begin{figure}[h]
\vspace{-5.5mm}
\centering
\hspace{-6mm}\subfloat[C-ILP vs. C-ALNS ]{\includegraphics[width = 4.5cm]{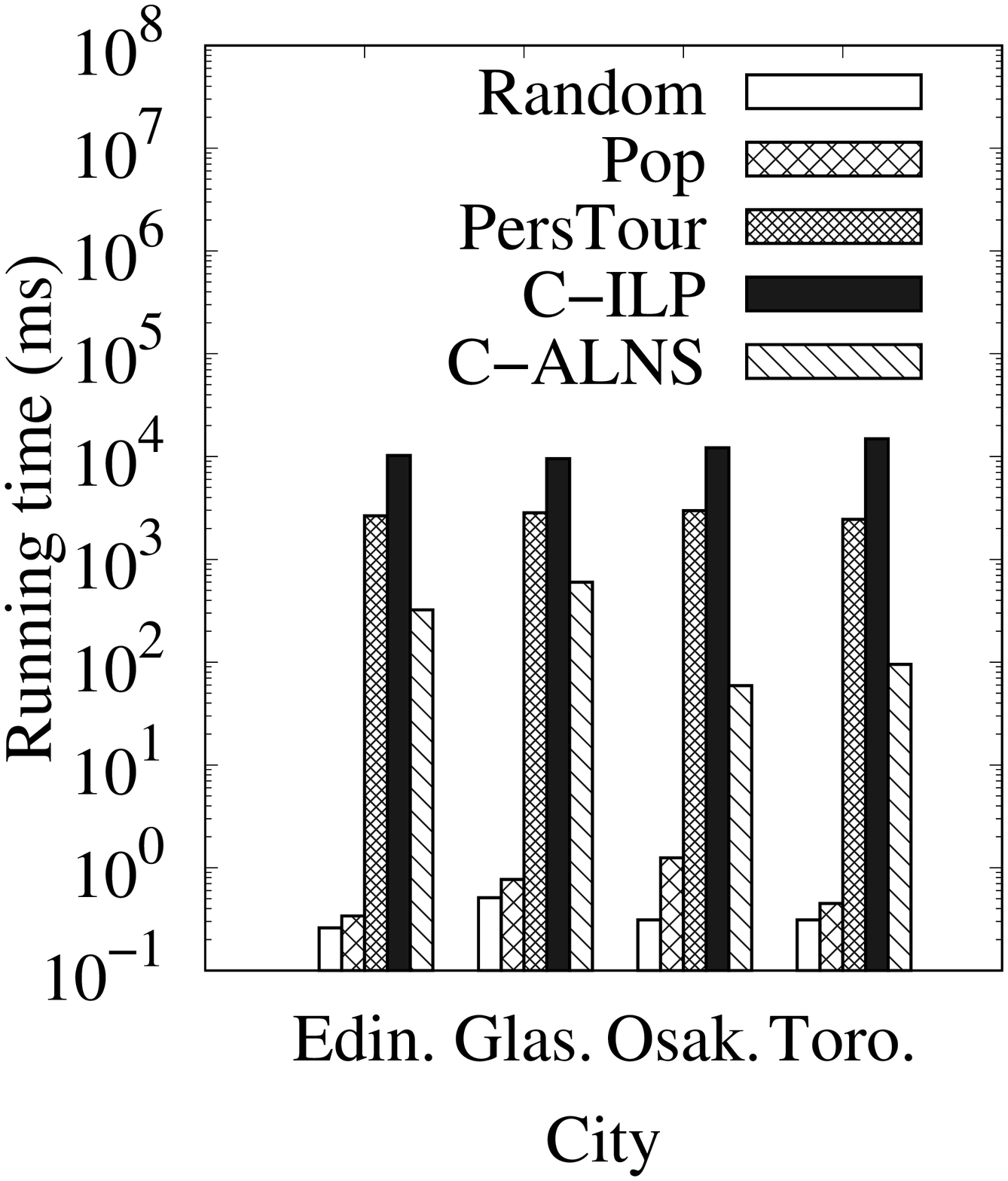}\label{fig:runtime}}\hspace{-8mm}
\subfloat[Impact of factors]{\includegraphics[width = 4.5cm]{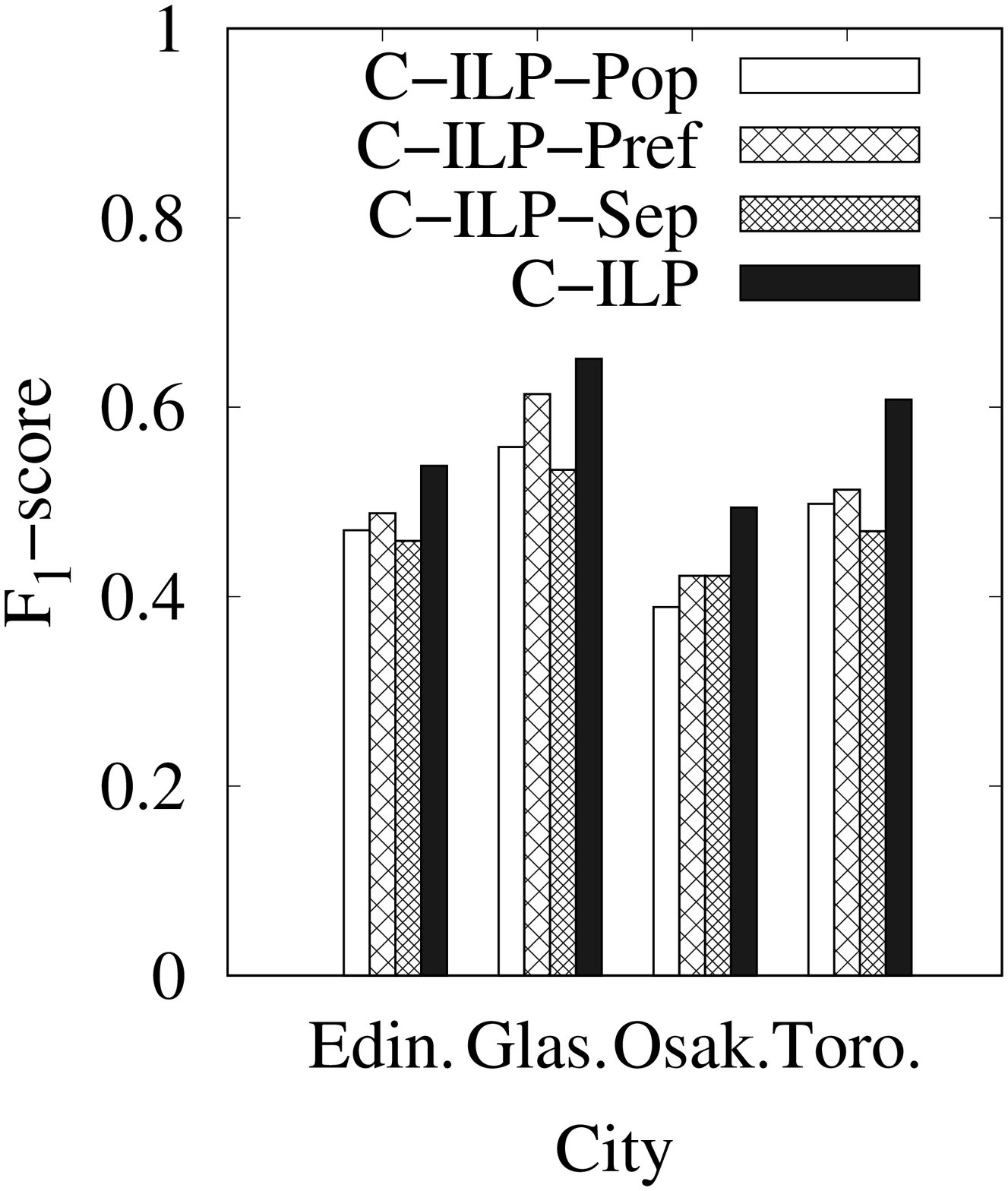}\label{fig:self}}\hspace{-4mm}
\vspace{-3mm}
\caption{Comparisons among proposed algorithms}
\label{fig:proposed_algorithms}
\vspace{-3mm}
\end{figure}

We compare the running times of C-ILP and C-ALNS in Figure~\ref{fig:runtime}. For completeness, we also include the running times of Random, Pop, and PersTour, but omit those of MF and POIRank as they resemble that of PersTour.
C-ALNS outperforms C-ILP and PersTour by orders of magnitude (note the logarithmic scale). The average running times of C-ILP and PersTour are $10^4$~ms  and $2.5\times 10^3$~ms, respectively, while that of C-ALNS is only around 300~ms, 600~ms, 60~ms, and 100~ms for the four datasets, respectively. Compared with C-ILP, C-ALNS reaches almost the same F$_1$-score while reducing the running time by up to $99.4\%$. Compared with PersTour, C-ALNS obtains up to $43\%$ improvement in F$_1$-score while reducing the running time by up to $97.6\%$. Random and Pop have the smallest running times but also very low trip quality as shown in Table~\ref{tab:result1}.

\emph{Impact of different factors.}
To investigate the contributions of POI popularities, user preferences, and co-occurring POIs in our embeddings, 
we implement two variants of C-ILP, namely, \textbf{C-ILP-Pop} and \textbf{C-ILP-Pref}. These two variants 
use POI embeddings that only learns POI popularities and jointly learns  POI popularities and user preferences, respectively. 
Fig.~\ref{fig:self} shows a comparison among C-ILP-Pop, C-ILP-Pref, and C-ILP. 
We see that the F$_1$-score increases as the POI embeddings incorporate more factors. 
This confirms the impact of the three factors. 
Moreover, we see that on the Edinburgh and Toronto datasets where POIs have more diverse POI co-occurrences (cf.~Section~\ref{sec:empirical}), 
the improvement of C-ILP (with co-occurring POIs in the embeddings)  over C-ILP-Pref is more significant. This demonstrates the effectiveness 
of our model to learn the POI co-occurrences. 
We also implement an algorithm that separately learns the impact of POI popularities, user preferences, and co-occurring POIs, denoted as \textbf{C-ILP-Sep}. 
The algorithm considers equal contribution of the three factors to recommend trips.
We see that C-ILP outperforms C-ILP-Sep consistently. This confirms the superiority of joint learning in our algorithm.

\begin{figure}[h]
\vspace{-5mm}
\centering
\hspace{-6mm}\subfloat[POI popularity]{\includegraphics[width = 4.5cm]{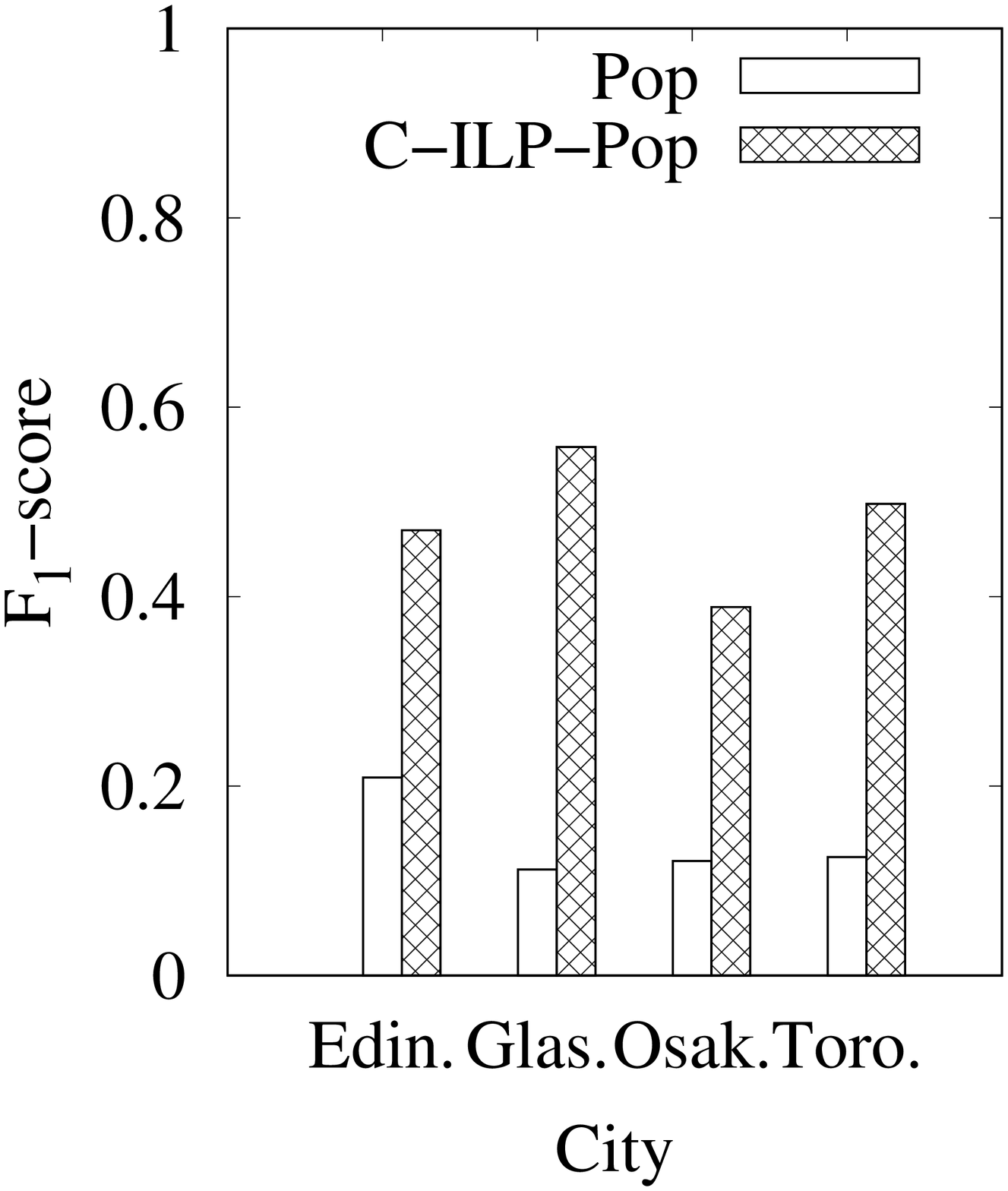}\label{fig:pop}}\hspace{-8mm}
\subfloat[User preferences]{\includegraphics[width = 4.5cm]{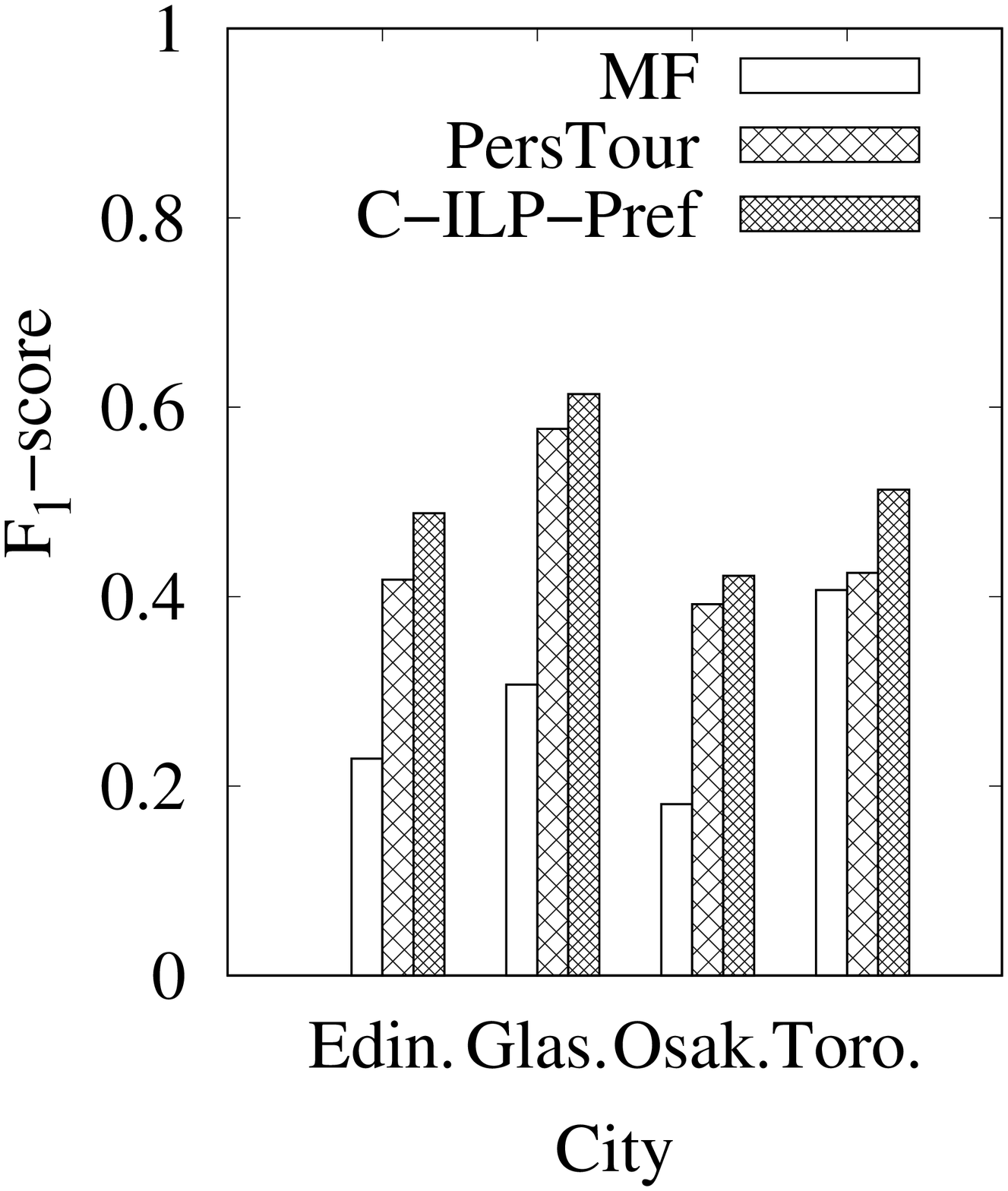}\label{fig:pref}}\hspace{-6mm}
\vspace{-3mm}
\caption{Impact of model learning capability}
\label{fig:model_capability}
\vspace{-3mm}
\end{figure}

\emph{Impact of model learning capability.}
To further show that our proposed POI embedding model has a better  learning capability, 
we compare C-ILP-Pop with Pop in Fig.~\ref{fig:pop}, since these two algorithms  only consider POI popularity. 
Similarly, we compare C-ILP-Pref with the baseline algorithms that considers user preferences, i.e., MF and PersTour, in Fig.~\ref{fig:pref}. 
In both figures, our models produce trips with higher F$_1$-scores, which confirms the higher learning capability of our models.

\section{Conclusions}
\label{sec:conclusions}
We proposed a  context-aware model for POI embedding. This model jointly learns the impact of POI popularities,  
co-occurring POIs, and user preferences over the probability of a POI being visited in a trip. 
To showcase the effectiveness of this model, we applied it to a trip recommendation problem named TripRec. 
We proposed two algorithms for TripRec based on the learned embeddings for both POIs and users. 
The first algorithm, C-ILP, finds the exact optimal trip by transforming and solving TripRec as an integer linear programming problem. 
The second algorithm, C-ALNS, finds a heuristically optimal trip but with a much higher efficiency based on the adaptive large neighborhood search technique. 
 We performed extensive experiments on real datasets. The results showed that the proposed algorithms using our 
 context-aware POI embeddings consistently outperform state-of-the-art algorithms in trip recommendation quality, 
 and the advantage is up to $43\%$ in F$_1$-score. C-ALNS reduces the running time for trip recommendation by $99.4\%$ 
comparing with C-ILP while retaining almost the same trip recommendation quality, i.e., only 0.2\% lower in F$_1$-score.


\balance


\bibliographystyle{abbrv}
\bibliography{ref}  


%
%
%
%

\end{document}